\documentclass[letterpaper,twocolumn,10pt,table]{article}
\usepackage{usenix2019_v3}
\usepackage{cite}
\usepackage{amsmath,amssymb,amsfonts}
\usepackage[linesnumbered,ruled,vlined,longend]{algorithm2e}
\SetKwComment{Comment}{$\triangleright$\ }{}
\usepackage{pifont}
\usepackage{graphicx}
\usepackage{textcomp}
\usepackage{xcolor}
\usepackage{tablefootnote}

\def\BibTeX{{\rm B\kern-.05em{\sc i\kern-.025em b}\kern-.08em
    T\kern-.1667em\lower.7ex\hbox{E}\kern-.125emX}}
\usepackage{pgfplots}
\usepackage{pgfplotstable}                      
\usetikzlibrary{pgfplots.statistics}        
\usetikzlibrary{shapes.symbols,shapes.geometric,shadows}
\usepackage{float}
\usepackage[english]{babel}
\usepackage[utf8]{inputenc}
\usepackage[T1]{fontenc}
\usepackage{amssymb}
\usepackage{amsmath}
\usepackage{bm}
\usepackage{amsfonts}
\usepackage{mathtools}
\usepackage{subcaption}
\usepackage{listings}
\usepackage{multirow}
\usepackage{array}
\usepackage{arydshln}
\usepackage[framemethod=tikz]{mdframed}
\usepackage[inline]{enumitem}
\usepackage{tikz}
\usetikzlibrary{arrows,chains,shapes,snakes,graphs,fit, positioning}

\usepackage{booktabs}
\usepackage{adjustbox}
\usepackage[disable]{todonotes}

\usepackage{wasysym}
\usepackage{caption}
\usepackage{titlesec}
\titlespacing*{\section}{0pt}{3.1ex plus 0.5ex minus .2ex}{0.9ex plus .2ex}
\titlespacing*{\subsection}{0pt}{2.7ex plus 0.5ex minus .2ex}{0.9ex plus .2ex}
\titlespacing*{\subsubsection}{0pt}{2.5ex plus 0.5ex minus .2ex}{1ex plus .2ex}
\titlespacing*{\paragraph}{0pt}{2.5ex plus 0.5ex minus .2ex}{1em}
\titlespacing*{\subparagraph} {\parindent}{2.5ex plus 0.5ex minus .2ex}{1em}

\usepackage{etoolbox}
\makeatletter
\patchcmd{\ttlh@hang}{\parindent\z@}{\parindent\z@\leavevmode}{}{}
\patchcmd{\ttlh@hang}{\noindent}{}{}{}
\makeatother

\usepackage{xspace} 
\definecolor{darkspringgreen}{rgb}{0.09, 0.45, 0.27}
\definecolor{tropicalrainforest}{rgb}{0.0, 0.46, 0.37}
\definecolor{persimmon}{rgb}{0.93, 0.35, 0.0}

\newcommand{\raidcomment}[1]{}  

\newcommand{\ntestedcodes}{6} 
\newcommand{\nbugs}{30}   
\newcommand{\nuafbugs}{11}  
\newcommand{\ncve}{7}       
\newcommand{\nfixbugs}{17}   

\newcommand{\benchncode}{11}         
\newcommand{\benchnbug}{13}         

\newcommand{\fullbenchncode}{17}         
\newcommand{\fullbenchnbug}{30}     

\usepackage{amsthm}
\theoremstyle{definition}
\newtheorem{definition}{Definition}

\definecolor{forestgreen}{RGB}{22,139,22}
\definecolor{feedback}{RGB}{110,017,178}
\definecolor{inputs}{RGB}{009,178,078}
\definecolor{sourcefile}{RGB}{255,170,025}

\definecolor{light-gray}{gray}{0.9}

\tikzstyle{rect} = [draw, rectangle, minimum height = 0.7cm, minimum width =
1.5cm]
\tikzstyle{leg} = [font=\it \footnotesize]
\tikzstyle{n} = [draw, ellipse, inner sep=2pt, font=\tiny]
\tikzstyle{l} = [font=\tiny, left]
\tikzstyle{txt} = [text width = 2.1cm, font=\scriptsize]
\tikzstyle{more} = [txt, text width = 4cm, text centered]

\newcommand{\uaf}{{\sc UAF}}
\newcommand{\afl}{{\sc AFL}}
\newcommand{\aflgo}{{\sc AFLGo}}
\newcommand{\aflgof}{{\sc AFLGo$_{F}$}}

\newcommand{\aflgob}{{\sc AFLGoB}}

\newcommand{\aflgobss}{{\sc AFLGoB}--ss}
\newcommand{\aflgobds}{{\sc AFLGoB}--ds}

\newcommand{\he}{{\sc Hawkeye}}
\newcommand{\heb}{{\sc HawkeyeB}}
\newcommand{\aflqemu}{{\sc AFL-QEMU}}
\newcommand{\gueb}{{\sc GUEB}}
\newcommand{\binsec}{{\sc Binsec}}
\newcommand{\ida}{{\sc IDA Pro}}

\newcommand{\uafuzz}{{\sc UAFuzz}}
\newcommand{\valgrind}{{\sc Valgrind}}
\newcommand{\oss}{{\sc OSS-Fuzz}}
\newcommand{\libfuzzer}{{\sc libFuzzer}}

\newcommand{\au}{$\hat A_{\mathrm{12U}}$}
\newcommand{\aaa}{$\hat A_{\mathrm{12A}}$}

\usepackage{filecontents}

\usepackage{stfloats}
\usepackage{cleveref}
\crefname{figure}{Figure}{Figures}
\crefname{section}{Section}{Sections}
\crefname{table}{Table}{Tables}
\crefname{algorithm}{Algorithm}{Algorithms}
\crefname{listing}{Listing}{Listings}
\crefname{appendix}{Appendix}{Appendixes}

\usepackage{tcolorbox}
\definecolor{mycolor}{rgb}{0.122, 0.435, 0.698}
\definecolor{gray1}{gray}{0.3}

\newcommand{\result}[1]{%
\begin{tcolorbox}[colframe=gray1,colback=gray!30,boxrule=0.5pt,arc=4pt,left=6pt,right=6pt,top=6pt,bottom=6pt,boxsep=0pt,width=\linewidth]{#1}
\end{tcolorbox}%
}

\renewcommand{\paragraph}[1]{\smallskip \noindent {\bf #1.}}

\definecolor{officegreen}{rgb}{0.0, 0.5, 0.0}
\usepackage{amssymb}
\usepackage{pifont}
\newcommand{\yes}{{\color{officegreen}\ding{51}}}%
\newcommand{\no}{{\color{red}\ding{55}}}%

\usepackage{listings}

\lstdefinelanguage{customC}{
	belowcaptionskip=1\baselineskip,
	tabsize=2,
	breaklines=true,
	xleftmargin=\parindent,
	language=C,
	showstringspaces=false,
	basicstyle=\scriptsize\ttfamily\color{black},
	keywordstyle=\bfseries\color{officegreen},
	commentstyle=\itshape\color{blue},
	stringstyle=\color{red},
	numbers=left,
	numbersep=5pt,
	numberstyle=\tiny\color{black},
	escapechar=@
}

\usepackage{siunitx}
\usepackage{tikz} 
\usepackage{pgfplots}
\usetikzlibrary{patterns}

\crefformat{subsection}{\S#2#1#3}
\crefformat{subsubsection}{\S#2#1#3}

\usetikzlibrary{decorations.pathreplacing,shapes,arrows,positioning}
\usetikzlibrary{calc} 
\usepackage[]{xcolor}
\usetikzlibrary{positioning}

\pgfplotsset{compat=newest} 
\usepgfplotslibrary{units} 

\pgfdeclarepatternformonly{mydots}{\pgfqpoint{-1pt}{-1pt}}{\pgfqpoint{2pt}{2pt}}{\pgfqpoint{2pt}{2pt}}%
{
	\pgfpathcircle{\pgfqpoint{0pt}{0pt}}{.3pt}
	\pgfpathcircle{\pgfqpoint{2pt}{2pt}}{.3pt}
	\pgfusepath{fill}
}

\author{
	{\rm Manh-Dung Nguyen}\\
	{\normalsize Univ. Paris-Saclay, CEA LIST, France}\\
	{\normalsize manh-dung.nguyen@cea.fr}
	\and
	{\rm S\'ebastien Bardin}\\
	{\normalsize Univ. Paris-Saclay, CEA LIST, France}\\
	{\normalsize sebastien.bardin@cea.fr}
	\and
	{\rm Richard Bonichon}\\
	{\normalsize Tweag I/O, France}\\
	{\normalsize richard.bonichon@tweag.io}
	\and
	{\rm Roland Groz}\\
	{\normalsize Univ. Grenoble Alpes, France}\\
	{\normalsize roland.groz@univ-grenoble-alpes.fr}
	\and
	{\rm Matthieu Lemerre}\\
	{\normalsize Univ. Paris-Saclay, CEA LIST, France}\\
	{\normalsize matthieu.lemerre@cea.fr}
} 

\begin{document}
	
\title{\vspace{-2em}Binary-level Directed Fuzzing for Use-After-Free Vulnerabilities}

\maketitle

\begin{abstract}
  {\it Directed  fuzzing}  focuses on automatically  testing  specific parts of the code    by taking advantage of additional information  such as (partial) bug stack trace,  patches  
or risky operations. Key applications include bug reproduction, patch testing and static analysis report verification. Although directed fuzzing has received 
a lot of attention recently, 
  hard-to-detect vulnerabilities such as Use-After-Free (\uaf) are still not well
addressed, 
especially at the binary level. 
We propose  \uafuzz, the first  (binary-level) directed greybox fuzzer dedicated to  
\uaf\ bugs. 
The technique features a fuzzing engine tailored to \uaf\ specifics, a lightweight code instrumentation and an efficient bug triage step. 
Experimental evaluation for bug reproduction on real cases 
demonstrates  that \uafuzz\ significantly outperforms  state-of-the-art  directed fuzzers in terms of fault detection rate, 
time to exposure and bug triaging.
\uafuzz\ has also  been proven  effective in patch testing, leading to the discovery of \nbugs\ new bugs (\ncve\ CVEs)  
in programs such as Perl, GPAC and GNU Patch.  
Finally, we provide to the community a large fuzzing benchmark dedicated to \uaf, built on both real codes and real bugs.     
\end{abstract}

\section{Introduction}
\paragraph{Context}
Finding bugs early is crucial
in the vulnerability management process. The recent rise of fuzzing \cite{miller1990empirical,manes2019art} in both
academia and industry, such as Microsoft's Springfield~\cite{springfield} and
Google's \oss~\cite{oss}, shows its ability to find various types of
bugs in real-world applications. {\it Coverage-based Greybox Fuzzing} (CGF), such as
\afl~\cite{afl} and \libfuzzer~\cite{libfuzzer}, leverages code coverage
information in order to guide input generation toward new parts of the program under test (PUT), exploring as many  program states as possible  in the hope of triggering
crashes. 
On the other hand, {\it Directed 
Greybox Fuzzing} (DGF)~\cite{bohme2017directed,chen2018hawkeye} aims to perform stress testing on pre-selected potentially vulnerable target locations, 
  with applications to different  security contexts:  
(1) \emph{bug reproduction}~\cite{jin2012bugredux,pham2015hercules,bohme2017directed,chen2018hawkeye}, 
(2) \emph{patch testing}~\cite{marinescu2013katch,peng20191dvul,bohme2017directed} 
or (3) static analysis report verification~\cite{christakis2016guiding,liang2019sequence}. 
Depending on the application, target locations are originated from {\it  bug stack traces}, patches or static analysis reports. 

We focus mainly on {\it bug reproduction}, which  is the most common practical application of DGF~\cite{jin2012bugredux,bohme2017directed,you2017semfuzz,chen2018hawkeye,liang2019sequence}. 
 It consists in  generating Proof-of-Concept  (PoC) inputs  of disclosed
 vulnerabilities given bug report information.  It is especially needed since
 only 54.9\% of usual bug reports can be reproduced due to missing information
 and users' privacy violation~\cite{217567}. Even with a PoC provided in the bug
 report, developers may still need to consider all corner cases of the bug in
 order to avoid regression bugs or incomplete fixes. In this
 case, providing more bug-triggering inputs becomes important  
 to facilitate  and accelerate 
 the repair process.   {\it Bug stack traces}, sequences of 
 function calls at the time a bug is triggered, are  widely used for guiding
 directed fuzzers~\cite{jin2012bugredux,bohme2017directed,chen2018hawkeye,liang2019sequence}. 
 Running a code on a PoC input under 
 profiling tools like AddressSanitizer
 (ASan)~\cite{serebryany2012addresssanitizer} or
 \valgrind~\cite{nethercote2007valgrind} will output such a bug stack trace.  

\paragraph{Problem} 
Despite tremendous progress in the past few years \cite{lafintel,li2017steelix,lemieux2018fairfuzz,aschermannredqueen,stephens2016driller,rawat2017vuzzer,angora,bohme2017directed,chen2018hawkeye,blazytko2019grimoire,you2019profuzzer,pham2019smart,fioraldi2019weizz}, 
current (directed or not) greybox fuzzers still have a hard time finding \textit{complex
  vulnerabilities} such as Use-After-Free (\uaf), non-interference or flaky
bugs~\cite{Bohme:2019:AST:3339132.3339136}, which 
require bug-triggering paths satisfying very specific properties. For example, \oss~\cite{oss,ossfuzzfive} or recent greybox fuzzers~\cite{bohme2017directed,rawat2017vuzzer,you2019profuzzer} only found a small number of \uaf. 
Actually, \textsc{Rode0day}~\cite{rode0day}, a continuous bug finding competition,
recognizes that fuzzers should aim to cover new bug classes like
\uaf\ in the future~\cite{fasano2019rode0day}, moving further from 
the widely-used LAVA~\cite{dolan2016lava} bug corpora  which only contains
buffer overflows. 

{\it We focus on UAF bugs}. They 
appear when a heap element is used after having been freed. 
The numbers of UAF bugs has increased in the National Vulnerability
Database (NVD)~\cite{nvd}. They are currently identified as {\it one of the most critical
exploitable vulnerabilities} due to the lack of mitigation techniques compared to
other types of bugs, and they may have serious consequences  such as data
corruption, information leaks and denial-of-service attacks.

\paragraph{Goal and challenges} {\it We focus on the problem of designing an efficient directed fuzzing method 
tailored for UAF.}  The technique must also be able to work at binary-level (no source-level instrumentation), as   source codes of 
 security-critical programs are not always available or may rely partly on third-party libraries.  
However, fuzzers targeting the detection of \uaf\ bugs confront themselves with the following challenges. 

\begin{enumerate}[label={\bf C\arabic*.},ref={\bf C\arabic*.},nosep]

\item~{\bf Complexity} -- Exercising \uaf\ bugs require to generate inputs triggering a \emph{
    sequence} of 3 events -- \textit{alloc}, \textit{free} and \textit{use} -- \emph{on the same memory location}, 
  spanning multiple functions of the PUT, where buffer overflows only require
  a single out-of-bound memory access. This combination of both {\it temporal} and {\it spatial} constraints 
  is  extremely difficult to meet in practice; 

\item~{\bf Silence} --  \uaf\ bugs often have \emph{no observable effect}, such as segmentation
  faults. In this case, fuzzers simply observing crashing behaviors do not
  detect that a test case triggered such a memory bug.  Sadly, popular profiling tools such as  ASan
  or \valgrind\ cannot be used in a fuzzing context  due to their high  runtime overhead. 
\end{enumerate}

\noindent Actually, current state-of-the-art directed fuzzers, namely \aflgo~\cite{bohme2017directed} and \he~\cite{chen2018hawkeye}, 
fail to address these challenges. 
First,  they are too generic and  therefore  do not cope   with the specificities of \uaf\ such as temporality -- their guidance metrics 
do not consider any notion of sequenceness.  
Second, they are completely blind to \uaf\ bugs, requiring to send all the many generated seeds to a profiling tool for an expensive extra check. 
Finally, current implementations of source-based DGF fuzzers typically suffer from  an expensive instrumentation step~\cite{aflgoissues}, 
e.g., \aflgo\ spent nearly $2h$ compiling and instrumenting 
 \texttt{cxxfilt} (Binutils).

{\centering\begin{table}[t]
	\centering \scriptsize
	\caption{Summary of existing greybox fuzzing techniques.}
	\label{tab:rw_small}
	\resizebox{\linewidth}{!}{%
		\begin{tabular}{l|ccc|c}
& \afl & \aflgo & \he & \uafuzz
\\
\hline
Directed fuzzing approach&\no&\yes&\yes&\yes \\
Support binary&\yes&\no&\no&\yes \\
UAF bugs oriented&\no&\no&\no&\yes \\
Fast instrumentation&\yes&\no&\no&\yes \\
UAF bugs triage&\no&\no&\no&\yes \\
\hline
\end{tabular} 
}
\end{table}}

\paragraph{Proposal}  We propose \uafuzz, {\it the first (binary-level) directed greybox fuzzer tailored to UAF bugs.}   
A quick comparison of \uafuzz\ with existing greybox fuzzers in terms of \uaf\ is presented in  \cref{tab:rw_small}. 
While we follow mostly the generic scheme of directed fuzzing, we carefully tune several of its key components to the specifics of \uaf:    

\begin{itemize}[nosep]

\item the {\it distance  metric}  favors shorter call chains leading to the target functions
that are more likely to include both allocation and free functions -- where sota directed fuzzers  rely on a generic CFG-based distance; 

\item {\it seed selection} is now based on a {\it sequenceness-aware target similarity metric} -- where sota directed fuzzers rely at best on  target coverage;  

\item our {\it power schedule} benefits from these new metrics, plus another one called {\it cut-edges} favoring prefix paths more likely to reach the whole target.    

\end{itemize}

\noindent Finally, the  bug triaging step piggy-backs on our previous metrics to 
pre-identifies  seeds as likely-bugs or not,  sparing a huge amount of queries  to the profiling tool for confirmation (\valgrind\ in our implementation).

\paragraph{Contributions} Our contribution is the following:
\begin{itemize}[nosep]

\item We design the first directed greybox fuzzing technique {\it tailored to} \uaf\
  bugs \raidcomment{working directly on executables} (\cref{sec:method}). Especially, we systematically revisit the three main ingredients of 
  directed fuzzing (selection heuristic, power schedule, input metrics) and specialize them to \uaf. 
These improvements are proven beneficial and complementary;

\item We develop a toolchain~\cite{uafuzz}
 on top of the state-of-the-art
  greybox fuzzer \afl\ \cite{afl} and the binary analysis platform \binsec
  \cite{binsec}, named \uafuzz, implementing the above method for \uaf\ directed fuzzing over binary codes (\cref{sec:implem}) and 
enjoying small overhead;

\item We construct and openly release~\cite{uafuzzbench}
the largest fuzzing benchmark  dedicated to \uaf, comprising  \textbf{\fullbenchnbug} real bugs from \textbf{\fullbenchncode} widely-used projects (including the few previous \uaf\ bugs found by directed fuzzers), in the hope of facilitating future \uaf\ fuzzing evaluation (\cref{app:sec:benchmark});

\item We evaluate our technique and tool in a bug reproduction setting
  (\cref{sec:experiment}), demonstrating   that \uafuzz\ is highly effective and significantly outperforms  state-of-the-art  competitors:   
2$\times$ faster in average to trigger bugs (up to 43$\times$), +34\% more successful runs in average (up to +300\%)  and 17$\times$ faster in triaging bugs 
(up to 130$\times$, with 99\% spare checks);  

\item Finally, \uafuzz\ is also   proven effective in patch testing (\cref{sec:zero}), leading to the discovery of \nbugs\ unknown bugs (11 UAFs, \ncve\ CVEs)
in projects like Perl, GPAC, MuPDF and  GNU Patch (including 4 buggy patches). 
So far, \nfixbugs\  have been fixed.   
\end{itemize}
\smallskip 

\noindent \uafuzz\ is the {\it first} directed greybox fuzzing approach tailored to detecting \uaf\ vulnerabilities (in binary) given only bug stack traces.
\uafuzz\ outperforms existing directed fuzzers on this class of vulnerabilities for bug reproduction and    
encouraging results have been obtained as well on   patch testing. 
We believe that our approach may also  be useful in slightly related contexts, for example  partial bug reports from static analysis or  other classes of vulnerabilities.

\section{Background}\label{sec:bg}

Let us first clarify  some notions used along the paper.

\subsection{Use-After-Free}
\label{sec:use-after-free}

\paragraph{Execution} An \emph{execution} is the complete sequence of
states executed by the program on an {\it input}.
An execution trace \emph{crashes} when it ends with a visible
error. The standard goal of  fuzzers is to find inputs  leading to crashes, as
 crashes are the first step toward exploitable vulnerabilities.

\paragraph{UAF bugs} 
 \emph{Use-After-Free} (\uaf) bugs happen  when dereferencing a
pointer to a heap-allocated object which is no longer valid (i.e., the pointer is
\emph{dangling}). Note that \emph{Double-Free} (DF) is a special case. 

\paragraph{UAF-triggering conditions} 
Triggering
a \uaf{} bug requires to find an input whose execution  covers  in
sequence \emph{three \uaf{} events}: an allocation (\emph{alloc}), 
a \emph{free}  and a \emph{use}  (typically, a dereference), {\it all three referring
to the same memory object}, as shown in~\cref{lst:uaf_ex}.

\begin{lstlisting}[language=customC,caption={Code snippet illustrating a UAF bug.},captionpos=b,label={lst:uaf_ex}]
char *buf = (char *) malloc(BUF_SIZE);
free(buf); // pointer buf becomes dangling
...
strncpy(buf, argv[1], BUF_SIZE-1); // Use-After-Free
\end{lstlisting}

Furthermore, this last \emph{use}  generally does not make the
execution immediately crash, as  a memory violation crashes a process
only when it accesses an address outside of the address space of the
process, which is unlikely with a dangling pointer.   
Thus, \uaf{} bugs go often unnoticed and are a good vector of  exploitation~\cite{younan2015freesentry,lee2015preventing}. 

\subsection{Stack Traces and Bug Traces}  \label{sec:bugtrace}
By inspection of the state of a process we can extract a \emph{stack
  trace}, i.e. the list of the function calls active in that
state. Stack traces are easily obtained from a process when it
crashes. As they provide (partial) information about the sequence of
program locations leading to a crash, they are extremely  valuable
for bug reproduction \cite{jin2012bugredux,bohme2017directed,chen2018hawkeye,liang2019sequence}. 

Yet, as crashes caused by \uaf{} bugs may happen 
long after the \uaf{} happened,  standard  stack traces usually do  not help in reproducing  \uaf\ bugs. Hopefully, profiling tools
for dynamically detecting memory corruptions, such as ASan~\cite{serebryany2012addresssanitizer} or \textsc{Valgrind}~\cite{nethercote2007valgrind},  
record the stack traces of {\it all memory-related events}:  when they detect
that an object is used after being freed, they actually report {\it three
stack traces}  (when the object is allocated, when it is freed  and when it is used
after being freed). 
\textit{We call such a sequence of three stack traces a
\textbf{(UAF) bug trace}}. When we use a bug trace as an input to try
to reproduce the bug, we call such a bug trace a \emph{target}.

{\centering\begin{figure}[ht!]\ttfamily\scriptsize
	\begin{tabular}{@{}l@{}}
\qquad // stack trace for the bad Use\\[-0.5pt]
==4440== Invalid read of size 1\\[-0.5pt]
==4440==    at 0x40A8383: vfprintf (vfprintf.c:1632)\\[-0.5pt]
==4440==    by 0x40A8670: buffered\_vfprintf (vfprintf.c:2320)\\[-0.5pt]
==4440==    by 0x40A62D0: vfprintf (vfprintf.c:1293)\\[-0.5pt]
==4440==    by 0x80AA58A: error (elfcomm.c:43)\\[-0.5pt]
==4440==    by 0x8085384: process\_archive (readelf.c:19063)\\[-0.5pt]
==4440==    by 0x8085A57: process\_file (readelf.c:19242)\\[-0.5pt]
==4440==    by 0x8085C6E: main (readelf.c:19318)\\[-0.5pt]
\\[-2pt]
\qquad // stack trace for the Free\\[-0.5pt]
==4440==  Address 0x421fdc8 is 0 bytes inside a block of size 86 free'd\\[-0.5pt]
==4440==    at 0x402D358: free (in vgpreload\_memcheck-x86-linux.so)\\[-0.5pt]
==4440==    by 0x80857B4: process\_archive (readelf.c:19178)\\[-0.5pt]
==4440==    by 0x8085A57: process\_file (readelf.c:19242)\\[-0.5pt]
==4440==    by 0x8085C6E: main (readelf.c:19318)\\[-0.5pt]
\\[-2pt]
\qquad // stack trace for the Alloc\\[-0.5pt]
==4440==  Block was alloc'd at\\[-0.5pt]
==4440==    at 0x402C17C: malloc (in vgpreload\_memcheck-x86-linux.so)\\[-0.5pt]
==4440==    by 0x80AC687: make\_qualified\_name (elfcomm.c:906)\\[-0.5pt]
==4440==    by 0x80854BD: process\_archive (readelf.c:19089)\\[-0.5pt]
==4440==    by 0x8085A57: process\_file (readelf.c:19242)\\[-0.5pt]
==4440==    by 0x8085C6E: main (readelf.c:19318)\\[-2ex]
	\end{tabular}
	\caption{Bug trace of CVE-2018-20623 (UAF) produced by \valgrind.}\label{fig:CVE-2018-20623}
\end{figure}
\vspace*{-1em}
}

\subsection{Directed Greybox Fuzzing}
\label{sec:cgf} \label{sec:direct-greyb-fuzz}
Fuzzing~\cite{miller1990empirical,manes2019art} consists in stressing a code under test through massive input generation in order to find bugs. 
 Recent {\it coverage-based greybox fuzzers}  
(CGF)~\cite{afl,libfuzzer} rely on  {\it lightweight} program analysis to guide the search -- typically through  coverage-based feedback. 
Roughly speaking,  a  {\it seed} (input) is {\it favored} (selected) when it reaches under-explored parts of the code,  and such favored seeds are then {\it mutated} 
to create new seeds for the code to be executed on. 
CGF is geared toward covering code in the large, in the hope of finding unknown vulnerabilities. 
On the other hand, {\it directed greybox fuzzing} (DGF)~\cite{bohme2017directed,chen2018hawkeye}   aims at reaching a  {\it pre-identified  potentially buggy part of the code}
 from a {\it target} 
(e.g., patch, static analysis report),  as often and fast as possible. Directed fuzzers follow  the general principles and architecture as CGF, but adapt the key components to 
their  goal, essentially favoring seeds {\it ``closer''} to the target. 
Overall directed fuzzers\footnote{And coverage-based fuzzers.} are built upon three main steps: (1) \emph{instrumentation} (distance pre-computation), (2) \emph{fuzzing} (including seed selection, power schedule and seed mutation) and (3) \emph{triage}.   

The standard core algorithm of DGF is presented in  \cref{alg:dgf} (the parts we modify in \uafuzz\ are in gray). Given a program $P$, a set of initial seeds $S_{0}$ and a target  $T$, 
the algorithm outputs a set of bug-triggering inputs $S'$. The fuzzing queue $S$ is initialized with the initial seeds in $S_{0}$ (line~\ref{dgf:fq}). 
 
\begin{enumerate}[nosep]
\item DGF first performs a static analysis (e.g., {\it target distance computation} for each basic block) and insert the instrumentation for dynamic coverage or distance information (line~\ref{dgf:instr}); 

\item  The fuzzer then repeatedly mutates inputs $s$ chosen from the fuzzing queue $S$ (line~\ref{dgf:ss}) until a timeout is reached. An input is selected either if it is \emph{favored} (i.e., believed to be interesting)  or with a small probablity $\alpha$  (line~\ref{dgf:prob}). 
%
Subsequently, DGF assigns the {\it energy} (a.k.a, the number \textit{M} of mutants to be created) to the selected seed $s$ (line~\ref{dgf:ps}). 
%
Then, the fuzzer generates $M$ new inputs by randomly applying some predefined mutation operators on seed $s$ (line~\ref{dgf:mutate}) and monitors their executions (line~\ref{dgf:run}). 
%
If the generated mutant $s'$ crashes the program, it is added to the set $S'$ of crashing inputs (line~\ref{dgf:crash}). Also,  newly generated mutants are added to the fuzzing queue\footnote{This is a general view. In practice, seeds regarded as very uninteresting are already discarded at this point.} (line~\ref{dgf:add});  

\item Finally, DGF returns $S'$ as the set of bug-triggering inputs (triage does nothing in standard DGF) (line~\ref{dgf:triage}).  

\end{enumerate}

{\centering\makeatletter
\newcommand{\removelatexerrorr}{\let\@latex@error\@gobble}
\makeatother
{\smallskip\removelatexerrorr%
\begin{algorithm}[htpb]\footnotesize
	\caption{Directed Greybox Fuzzing}
	\label{alg:dgf}
	
	\SetKwInOut{Input}{Input}
	\SetKwInOut{Output}{Output}

	\Input{ Program $P$; Initial seeds $S_{0}$; Target locations $T$ }
	\Output{ Bug-triggering seeds $S'$ }
	\BlankLine
	$S' := \emptyset;\ S :=  S_{0}$\Comment*[r]{$S$: the fuzzing queue}\label{dgf:fq}
	\hspace{-0.12cm}\colorbox{lightgray}{\parbox{\dimexpr\columnwidth-53\fboxsep\relax}{$P' \leftarrow \textsc{preprocess}(P,\ T)$}}\Comment*[r]{\textbf{phase 1}: Instrumentation}\label{dgf:instr}
	\While(\Comment*[f]{\textbf{phase 2}: Fuzzing}){timeout not exceeded}{
		\For{s $\in$ S}{\label{dgf:ss}
			\If{\colorbox{lightgray}{\parbox{\dimexpr\columnwidth-63.5\fboxsep\relax}{\textsc{is\_favored}($s$)}} or $rand() \leq \alpha$}{\Comment*[h]{seed selection, $\alpha$: small probability\label{dgf:prob}
				}\\
				\hspace{-0.12cm}\colorbox{lightgray}{\parbox{\dimexpr\columnwidth-51.0\fboxsep\relax}{$M := \textsc{assign\_energy}(s)$}}\Comment*[r]{power schedule}\label{dgf:ps}
				\For{i $\in$ \normalfont{1 ... M}}{
					$s' := \textbf{mutate\_input}(s)$\Comment*[r]{seed mutation}\label{dgf:mutate}
					$res := \textbf{run}(P',\ s',\ T)$;\label{dgf:run}\\
					\If{is\_crash(res)}{
						$S'\ :=\ S'\ \cup\ \{s'\}$\Comment*[r]{crashing inputs}\label{dgf:crash}
					} 
                                        \Else{$S\ :=\ S\ \cup\ \{s'\}$;\label{dgf:add}\\}

				}
			}
		}
	}
	\hspace{-0.12cm}\colorbox{lightgray}{\parbox{\dimexpr\columnwidth-58.5\fboxsep\relax}{$S'\ =\ \textsc{triage}(S,\ S')$}}\Comment*[r]{\textbf{phase 3}: Triage}\label{dgf:triage}
	\Return $S'$;
\end{algorithm}}
}

\aflgo~\cite{bohme2017directed} was the first to propose  a CFG-based distance to evaluate the proximity between a seed execution and multiple targets, together with  a simulated annealing-based power schedule. 
\he~\cite{chen2018hawkeye} keeps the CFG-based view but improves its accuracy\footnote{Possibly at the price of both higher pre-computation costs due to more precise static analysis and runtime overhead due to complex seed metrics.}, brings a seed selection heuristic partly based on target coverage (seen as a set of locations) and proposes adaptive mutations.
\section{Motivation}\label{sec:motiv}
The toy example in \cref{lst:motiv}  contains a \uaf\ bug due to a missing \texttt{exit()} call, a
common root cause in  such bugs (e.g., CVE-2014-9296, CVE-2015-7199). The
program reads a file and copies its contents into a buffer
\texttt{buf}. Specifically, a memory chunk pointed at by \texttt{p} is allocated
(line~\ref{cs:p_alloc}), then \texttt{p\_alias} and \texttt{p} become aliased 
(line~\ref{cs:alias}). The memory pointed by both pointers is freed in function
\texttt{bad\_func} (line~\ref{cs:free}). The \uaf\ bug occurs when the freed
memory is dereferenced again via \texttt{p} (line~\ref{cs:use}).

\paragraph{Bug-triggering conditions} The \uaf\ bug is triggered iff the first
three bytes of the input are `\texttt{AFU}'.  To quickly detect this bug,
fuzzers need to explore the right path through the \texttt{if} part of
conditional statements in lines \ref{cs:alloc_cond}, \ref{cs:free_cond} and
\ref{cs:use_cond} in order to cover in sequence the three \uaf\ events
\textit{alloc}, \textit{free} and \textit{use}
respectively. 
It is worth
noting that this \uaf\ bug does not make the program crash, hence existing
greybox fuzzers without sanitization will not detect this memory error.

\paragraph{Coverage-based Greybox Fuzzing}
Starting with an empty seed, \afl\ quickly generates 3 new inputs (e.g.,
`\texttt{AAAA}', `\texttt{FFFF}' and `\texttt{UUUU}') triggering individually the 3 
\uaf\ events. None of these seeds triggers the bug. 
As the probability of generating an input starting with `\texttt{AFU}` from an empty
seed  is extremely small, the  coverage-guided
mechanism is not effective here in tracking a sequence of \uaf\ events 
even though each individual event is easily triggered.

\paragraph{Directed Greybox Fuzzing} 
Given a bug trace (\ref{cs:palias_alloc} -- \textit{alloc},
\ref{cs:func}, \ref{cs:bad_func}, \ref{cs:motiv_free} -- \textit{free}, \ref{cs:use}
-- \textit{use}) generated for example by ASan, DGF prevents the fuzzer
from exploring undesirable paths, e.g., the \texttt{else} part at line \ref{cs:else} in
function \texttt{func}, in case the condition at line \ref{cs:free_cond} is more
complex. 
Still, directed fuzzers have their own blind spots. For example, standard DGF seed selection mechanisms favor 
a seed whose execution trace covers many locations in targets, instead of trying to reach these locations 
in a given order. For example, regarding a target 
$ (A,\ F,\ U)$, standard DGF distances \cite{bohme2017directed,chen2018hawkeye}  do  not discriminate between an input
$s_{1}$ covering a path $A\rightarrow F\rightarrow U$ and another input $s_{2}$
covering 
$U\rightarrow A\rightarrow F$. 
The lack of ordering in exploring target locations makes \uaf\ bug
detection very challenging for existing directed fuzzers. 
Another example:  
the power function proposed by \he~\cite{chen2018hawkeye} may assign much energy
to a seed whose trace  does not reach the target function, implying 
that it could get lost on the toy example in the \texttt{else} part at line \ref{cs:else} in
function \texttt{func}. 

\begin{lstlisting}[float,floatplacement=htpb,language=customC,caption={Motivating example.},captionpos=b,label={lst:motiv},belowskip=-3 \baselineskip]
int *p, *p_alias;
char buf[10];
void bad_func(int *p) {free(p);@\label{cs:motiv_free}@} /* exit() is missing */
void func() {
	if (buf[1] == 'F')@\label{cs:free_cond}@
		bad_func(p);@\label{cs:bad_func}@
	else /* lots more code ... */@\label{cs:else}@
}
int main (int argc, char *argv[]) {
	int f = open(argv[1], O_RDONLY);
	read(f, buf, 10);
	p = malloc(sizeof(int));@\label{cs:p_alloc}@
	if (buf[0] == 'A'){@\label{cs:alloc_cond}@
		p_alias = malloc(sizeof(int));@\label{cs:palias_alloc}@
		p = p_alias;@\label{cs:alias}@
	}
	func();@\label{cs:func}@
	if (buf[2] == 'U')@\label{cs:use_cond}@
		*p = 1;@\label{cs:use}@
	return 0;
}
\end{lstlisting}

\paragraph{A glimpse at \uafuzz} 
We  rely in particular  on 
modifying the seed selection heuristic w.r.t.
  the number of targets covered by an execution trace (\cref{sec:ss})
  and
bringing target ordering-aware seed metrics to DGF (\cref{sec:udm}). 

On the toy example, \uafuzz\ generates inputs to progress towards the expected target sequences. For
example, in the same fuzzing queue containing 4 inputs, the mutant
`\texttt{AFAA}', generated by mutating the seed `\texttt{AAAA}', is discarded by \afl\ as it does not increase code coverage. However, since it has maximum value of
target similarity metric score (i.e., 4 targets including lines \ref{cs:palias_alloc}, \ref{cs:func}, \ref{cs:bad_func}, \ref{cs:motiv_free}) compared to all 4 previous inputs in the queue (their scores are 0 or 2), this mutant is selected by \uafuzz\ for subsequent
fuzzing campaigns. By continuing to fuzz `\texttt{AFAA}', \uafuzz\ eventually
produces a bug-triggering input, e.g., `\texttt{AFUA}'.

\paragraph{Evaluation}
{\it \aflgo\ (source-level) cannot detect the UAF bug within 2 hours\footnote{\aflqemu\ did not succeed either.}\footnote{\he\ is not available and thus could not be tested.}, while \uafuzz\ (binary-level) is able to trigger it within  20 minutes. Also,  \uafuzz\  sends a  
 single 
input to \valgrind\  for confirmation (the right PoC input), while  \aflgo\ sends  120 inputs.}

 \section{The \uafuzz\ Approach}\label{sec:method}

\uafuzz\ is made out of several components encompassing seed selection
(\cref{sec:ss}), input metrics (\cref{sec:uaf-based-distance}),  power schedule
(\cref{sec:power}), and seed triage (\cref{sec:effic-analysis}). Before
detailing these aspects, let us start with an overview of the approach.
 
{\centering\begin{figure}[h]
	\centering
	\hspace*{-3em}
	\begin{adjustbox}{width=1.2\linewidth}
	\tikzstyle{block} = [rounded corners,rectangle, draw, text width=4em, text centered, inner sep=2pt, minimum height=3em]
	\tikzstyle{plain} = [text centered, text width=10em, inner sep=0pt, minimum height=3em]
	\tikzstyle{line} = [draw, -latex',very thick]

\begin{tikzpicture}[node distance = 1em, auto]
\node[inner sep=0pt] (bin) {\includegraphics[width=2em]{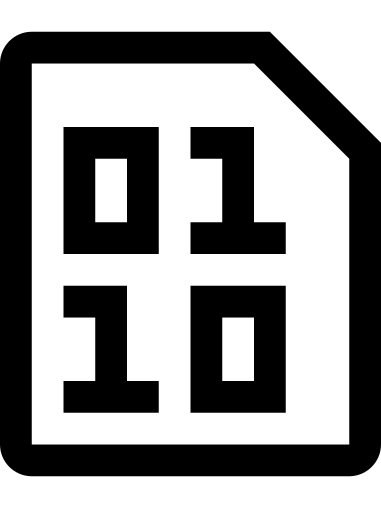}};
\node [plain] at (0,-0.7) (fuzz-lb) {Binary};
\node[inner sep=0pt,below=2em of bin] (target) {\includegraphics[width=2em]{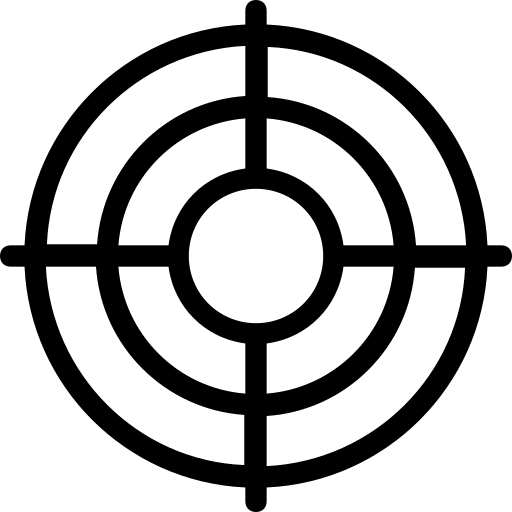}};
\node [plain] at (0,-2.1) (fuzz-lb) {Targets};

\node [block, very thick, node distance=1em, text width=3em, minimum height=5em] at ([xshift=4em,yshift=-2em]bin.east) (instr) {};
\node [draw,rounded corners,thick] at ([yshift=-1.5em]instr.north)(cg){CG};
\node [draw,rounded corners,thick] at ([yshift=1.5em]instr.south)(cfg){CFGs};
\node [plain, above of=instr,node distance=3.5em] (instr-lb) {Computation};

\node [block, very thick, node distance=1em, text width=9em, minimum height=9em] at ([xshift=6.5em]instr.east) (metrics) {};
\node [draw,rounded corners,thick] at ([yshift=-1.5em]metrics.north)(ds){UAF-based Distance};
\node [draw,rounded corners,thick] at (metrics.center)(cs){Cut-edge Coverage};
\node [draw,rounded corners,thick] at ([yshift=1.5em]metrics.south)(ts){Target Similarity};
\node [plain, above of=metrics,node distance=5.5em,] (metrics-lb) {Input Metrics};

\node [block,very thick] at ([xshift=4.5em]metrics.east)(ss){Seed Selection};
\node [block,very thick] at ([xshift=4.5em]ss.east)(ps){Power Schedule};
\node [block,very thick] at ([xshift=4.5em]ps.east)(triage){UAF Triage};

\node[inner sep=0pt] at ([xshift=2.5em]triage.east)(bug1) {\includegraphics[width=2em]{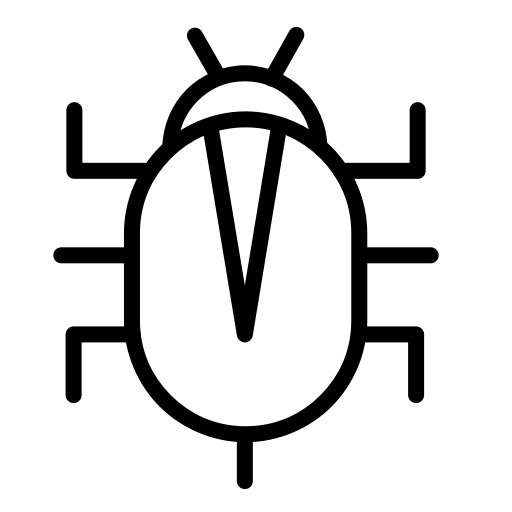}};
\node [plain] at (14.4,-1.3) (fuzz-lb) {UAF bugs};

\path [line] (bin) -- (instr);
\path [line] (target) -- (instr);
\path [line] (instr) -- (metrics);
\path [line] (metrics) -- (ss);
\path [line] (ss) -- (metrics);
\path [line] (ss) -- (ps);
\path [line] (ps) -- (ss);
\path [line] (ps) -- (triage);
\path [line] (triage) -- (bug1);

\coordinate (1) at (-0.5,-3.0);
\coordinate (2) at (2.7,-3.0);
\coordinate (3) at (11.3,-3.0);
\coordinate (4) at (14.8,-3.0);
\node [plain] at (1.1,-2.8) (instr-lb) {\textbf{Instrumentation}};
\node [plain] at (7.4,-2.8) (fuzz-lb) {\textbf{Fuzzing}};
\node [plain] at (13.2,-2.8) (triage-lb) {\textbf{Triage}};
\draw [|-,thick] (1)node[left]{} -- (1 -| 2)node[right](){};
\draw [|-,thick] (2)node[left]{} -- (2 -| 3)node[right]{};
\draw [|-|,thick] (3)node[left]{} -- (3 -| 4)node[right]{};

\end{tikzpicture}
	\end{adjustbox}
	\caption{Overview of \uafuzz.}\label{fig:uafuzz}
\end{figure}
}

We aim to find an
input fulfilling both control-flow (temporal) and runtime (spatial) conditions to trigger the \uaf\
bug. We solve this problem by bringing \uaf\ characteristics
into DGF in order to generate more potential inputs reaching targets in sequence
w.r.t. the \uaf\ expected bug trace. 
\cref{fig:uafuzz} depicts the general picture. Especially:  

\begin{itemize}[nosep]

\item We propose three dynamic seed metrics specialized for \uaf\
  vulnerabilities detection. The distance metric approximates how
  close a seed is to all target locations (\cref{sec:udm}), and takes
  into account the need for the seed execution trace to cover the
  three UAF events in order. The cut-edge coverage metric
  (\cref{sec:ccm}) measures the ability of a seed to take the correct
  decision at important decision nodes. Finally, the target
  similarity metric concretely assesses how many targets a seed
  execution trace covers at runtime (\cref{sec:tsm});

\item Our seed selection strategy (\cref{sec:ss}) favors seeds covering more
  targets at runtime.  The power scheduler determining the energy for each
  selected seed based on its metric scores during the fuzzing process is
  detailed in \cref{sec:power}; 

\item Finally, we take advantage of our previous metrics to pre-identify likely-PoC inputs that are sent to 
  a profiling tool (here \valgrind) for bug confirmation, avoiding many useless checks  
  (~\cref{sec:effic-analysis}). 
\end{itemize}

\subsection{Bug Trace Flattening}

{\centering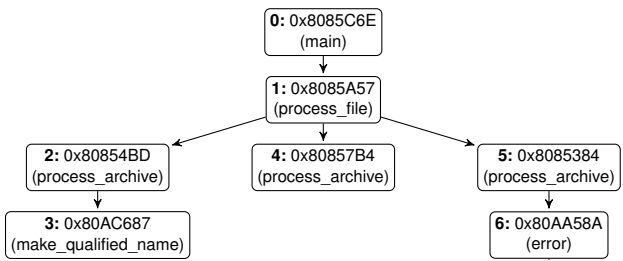
\begin{figure}[b]
	\centering
\begin{tikzpicture}[->,>=stealth',shorten >=1pt, xscale=2, yscale=0.6,
node/.style={inner sep=2pt,minimum height=1em,rectangle,rounded corners=2pt,draw,font=\sffamily\scriptsize},
]

\node[node,align=center] (0) at (0,0) [] {\textbf{0:} 0x8085C6E\\ (main)};
\node[node,align=center] (1) at (0,-1.5)  {\textbf{1:} 0x8085A57\\ (process\_file)};
\node[node,align=center] (2) at (-1.5,-3)  {\textbf{2:} 0x80854BD\\ (process\_archive)};
\node[node,align=center] (3) at (-1.5,-4.5)  {\textbf{3:} 0x80AC687\\ (make\_qualified\_name)};
\node[node,align=center] (4) at (0,-3)  {\textbf{4:} 0x80857B4\\ (process\_archive)};
\node[node,align=center] (5) at (1.5,-3)  {\textbf{5:} 0x8085384\\ (process\_archive)};
\node[node,align=center] (6) at (1.5,-4.5)  {\textbf{6:} 0x80AA58A\\ (error)};

\coordinate (after6) at (1.5,-5.5);

\draw[] (0) -- (1) node[]{};
\draw[] (1) -- (2) node[]{};
\draw[] (2) -- (3) node[]{};
\draw[] (1) -- (4) node[]{};
\draw[] (1) -- (5) node[]{};
\draw[] (5) -- (6) node[]{};
\draw[dotted] (6) -- (after6) node[]{};
\end{tikzpicture}\vspace{-3mm}
\caption{Reconstructed tree from CVE-2018-20623 (bug trace from Figure~\ref{fig:CVE-2018-20623}). The preorder traversal of this tree is simply  $0\rightarrow1\rightarrow2\rightarrow3(n_{alloc})\rightarrow4(n_{free})\rightarrow5\rightarrow6(n_{use})$.}\label{fig:dct}
\end{figure}

}

A bug trace (\cref{sec:bugtrace}) is a sequence of stack traces, i.e. it is a large  object 
not fit for the lightweight instrumentation required by greybox
fuzzing. The most valuable information that we need to extract from a
bug trace is the sequence of basic blocks (and functions) that were
traversed, which is an easier object to work with. We call this
extraction \emph{bug trace flattening}.

The operation works as follows. First, each of the three stack-traces
is seen as a path in a call tree; we thus merge all the stack traces
to re-create that tree. Some of the nodes in the tree have several children; we
make sure that the children are ordered according to the ordering of the UAF
events (i.e. the child coming from the \emph{alloc} event comes before
the child coming from the \emph{free} event). Figure~\ref{fig:dct} shows
an example of tree for the bug trace given in
Figure~\ref{fig:CVE-2018-20623}. 

Finally, we perform a preorder traversal of this tree to get a
sequence of target locations (and their associated functions), which we
will use in the following algorithms.

\subsection{Seed Selection based on Target Similarity}\label{sec:ss}

Fuzzers generate a large number of inputs so that smartly selecting the seed 
from the fuzzing queue to be mutated in the next fuzzing campaign
is crucial for effectiveness. Our seed
selection algorithm is based on two insights. First, we \emph{should
  prioritize seeds that are most similar to the target bug
  trace}, as the goal of a directed fuzzer is to find bugs covering
the target bug trace. Second, \emph{target similarity should take
  ordering (a.k.a.~sequenceness) into account}, as traces covering sequentially a number
of locations in the target bug trace are closer to the target than
traces covering the same locations in an arbitrary order.

\subsubsection{Seed Selection}

\begin{definition}{}
  A \textit{max-reaching input} is an input $s$ whose execution trace is the
  most similar to the target bug trace $T$ so far, where most similar means
  ``having the highest value as compared by a target similarity metric
  $t(s,T)$''.
\end{definition}
\vspace{-4mm}

{\centering\begin{algorithm}[htpb]\footnotesize
	\caption{\textsc{is\_favored}}
	\label{alg:ss}
	
	\SetKwInOut{Input}{Input}
	\SetKwInOut{Output}{Output}

	\Input{ A seed $s$ }
	\Output{ \textit{true} if $s$ is favored, otherwise \textit{false} }
	\BlankLine
	\textbf{global} $t_{max} = 0$\Comment*[r]{maximum target similar metric score}\label{ss:tmax}
	\lIf(\Comment*[f]{update $t_{max}$}\label{ss:uptmax}){$t(s) \geq t_{max}$}{\label{ss:newtmax}
		$t_{max} = t(s)$; \Return \textit{true}} 
	\lElseIf(\Comment*[f]{increase coverage}){$new\_cov(s)$}{\Return \textit{true}}
        \lElse{\Return \textit{false}}

\end{algorithm}
}%
\todo{rg: remove algo 2?}%
We mostly select and mutate max-reaching inputs during the fuzzing
process. Nevertheless, we also need to improve code coverage, thus
\uafuzz\ also selects inputs that cover new paths, with a small
probability $\alpha$ (\cref{alg:dgf}). In our experiments, the probability of selecting the
remaining inputs in the fuzzing queue that are less favored is $1\%$ like \afl~\cite{afl}.

\subsubsection{Target Similarity Metrics}\label{sec:tsm}

A \emph{target similarity metric} $t(s,T)$ measures the similarity
between the execution of a seed $s$ and the target \uaf{} bug trace
$T$. We define 4 such metrics, based on whether we consider ordering
of the covered targets in the bug trace ($P$), or not ($B$) -- $P$ stands
for Prefix, $B$ for Bag; and
whether we consider the full trace, or only the three UAF events
($3T$):

\begin{itemize}[noitemsep,nosep,leftmargin=4mm]
\item Target prefix $t_P(s,T)$: locations in $T$ covered in sequence by executing $s$ until first divergence; 
\item UAF prefix $t_{3TP}(s,T)$: UAF events of $T$ covered in sequence by executing $s$ until first divergence; 
\item \mbox{Target bag $t_B(s,T)$: locations in $T$ covered by executing $s$}; 
\item UAF bag $t_{3TB}(s,T)$: UAF events of $T$ covered by $s$.  \raidcomment{execu\-ting $s$.}
\end{itemize}

\noindent For example, using \cref{lst:motiv}, the 4 metric values of a seed $s$ `\texttt{ABUA}' w.r.t. the \uaf\ bug trace $T$ are:\\ $t_P(s,T) = 2$, $t_{3PT}(s,T) = 1$, $t_B(s,T) = 3$ and $t_{3TB}(s,T) = 2$.

\begin{figure}[htpb]
	\centering
	\begin{adjustbox}{width=0.7\linewidth}
		\tikzstyle{line} = [thin, draw, -latex']
		\begin{tikzpicture}[yscale=0.7]
		\node[inner sep=3pt] (p) at (1.5,0) {\footnotesize Prefix (\textit{P})};
		\node[left,text centered,inner sep=3pt] (3tp) at (0.6,-1) {\footnotesize UAF Prefix (\textit{3TP})};
		\node[right,text centered,inner sep=3pt] (b) at (2.7,-1) {\footnotesize Bag (\textit{B})};
		\node[inner sep=3pt] (3tb) at (1.5,-2.0) {\footnotesize UAF Bag (\textit{3TB})};
	
		\path [line] (p) -- (3,-0.7);
		\path [line] (p) -- (0,-0.7);
		\path [line] (3,-1.3) -- (3tb);
		\path [line] (0,-1.3) -- (3tb);
		
		\end{tikzpicture}
	\end{adjustbox}
	\caption{Precision lattice for Target Similarity Metrics}\label{fig:lattice}
\end{figure}
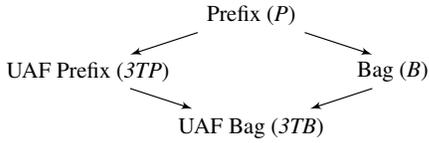

These 4 metrics have different degrees of \emph{precision}. A metric $t$ is
said \emph{more precise than} a metric $t'$ if, for
any two seeds $s_1$ and $s_2$: $t(s_1,T) \ge t(s_2,T) \;\Rightarrow\;
t'(s_1,T) \ge t'(s_2,T) $.
\cref{fig:lattice} compares our 4 metrics w.r.t their relative precision.

\noindent

\subsubsection{Combining Target Similarity Metrics}

Using a precise metric such as $P$ allows to better assess progression towards
the goal. In particular, $P$ can distinguish seeds that match the target bug
trace from those that do not, while other metrics cannot. 
On the other hand, a less precise metric provides information that precise
metrics do not have. For instance, $P$ does not measure any difference between
traces whose suffix would match the target bug trace, but who would diverge from
the target trace on the first locations (like `\texttt{UUU}' and `\texttt{UFU}' on \cref{lst:motiv}), while $B$ can.

To take benefit from both precise and imprecise metrics, we
combine them using a lexicographical order. Hence, the
\textit{P-3TP-B} metric is defined as:
\[ t_{P{-}3T\!P{-}B}(s,T) \triangleq  \langle t_{P}(s,T), t_{3TP}(s,T), t_{B}(s,T) \rangle  \]

This combination  favors first seeds that cover the most
locations in the prefix, then (in case of tie) those reaching the
most number of UAF events in sequence, and finally (in case of 
tie) those that reach the most locations in the target.
Based on  preliminary investigation, we default to 
\textit{P-3TP-B} for seed selection in \uafuzz.

\subsection{UAF-based Distance}\label{sec:udm} \label{sec:uaf-based-distance}
One of the main component of directed greybox fuzzers is the
computation of a \emph{seed distance}, which is an evaluation of a
distance from the execution trace of a seed $s$ to the
target. The main heuristic here is that if the execution trace of $s$ is close to the
target, then $s$ is close to an input that would cover the target, which makes $s$ an interesting seed.
In existing directed greybox fuzzers \cite{aflgo,chen2018hawkeye}, the seed
distance is computed to a target which is a single location or a set
of locations. This is not appropriate for the reproduction of UAF
bugs, that must go through 3 different locations in sequence. Thus, we
propose to modify the seed distance computation to take into account
the need to reach the locations in order.

\subsubsection{Zoom: Background on Seed Distances}
Existing directed greybox fuzzers \cite{aflgo,chen2018hawkeye} compute the
distance $d(s,T)$ from a seed $s$ to a target $T$ as follows.

\paragraph{\aflgo's seed distance \cite{aflgo}} 
The {\it seed distance} $d(s,T)$ is  
defined as the (arithmetic) mean of the {\it basic-block distances}  
 $d_b(m,T)$, for each basic block
$m$ in the execution trace of $s$. 

The {\it basic-block distance} $d_b(m,T)$ is defined using the length of the
intra-procedural shortest path from $m$ to the basic block of a
``call'' instruction, using the CFG of the function containing $m$;
and the length of the inter-procedural shortest path from the function
containing $m$ to the target functions $T_f$ (in our case, $T_f$ is
the function where the \emph{use} event happens), using the call graph.

\paragraph{\he{}'s enhancement \cite{chen2018hawkeye}} The main factor in this  seed distance
computation is computing distance between functions in the call
graph. To compute this, \aflgo{} uses the original call graph with
every edge having weight 1. \he{} improves this computation by
proposing the augmented adjacent-function distance (AAFD), which
changes the edge weight from a caller function $f_a$ and a callee
$f_b$ to $w_{Hawkeye}(f_a,f_b)$. The idea is to favor edges in the
call graph where the callee can be called in a variety of situations,
i.e. appear several times at different locations.

\subsubsection{Our UAF-based Seed Distance}\label{sec:usd}
Previous seed distances  \cite{aflgo,chen2018hawkeye}  do not account for any order among the target locations, while it is 
essential for \uaf.   
We address this issue by  modifying the distance between functions in the call graph to favor paths
that {\it sequentially} go through the three UAF events \emph{alloc}, \emph{free} and
\emph{use} of the bug trace. This is done by decreasing the weight of the edges
in the call graph that are likely to be between these events, using 
lightweight static analysis.

This analysis first computes $R_{alloc}, R_{free},$ and $R_{use}$,
i.e., the sets of functions that can call respectively the
\emph{alloc}, \emph{free}, or \emph{use} function in the bug trace --
the \emph{use} function is the one where the \emph{use} event
happens.
Then, we consider each call edge between $f_a$ and $f_b$ as indicating
a direction: either downward ($f_a$ executes, then calls $f_b$), or
upward ($f_b$ is called, then $f_a$ is executed). Using this we compute,
for each direction, how many events in sequence can be covered by
going through the edge in that direction. For instance, if $f_a \in R_{alloc}$ and
$f_b \in R_{free} \cap R_{use}$, then taking the $f_a \to f_b$ call
edge possibly allows to cover the three UAF events in sequence. To
find double free, we also include, in this computation, call edges
that allow to reach two \emph{free} events in sequence. 

Then, we favor a call edge from $f_a$ to $f_b$  by
decreasing its weight, based on how
many events in sequence the edge allows to cover. In our experiments, we use
the following $\Theta_{UAF}(f_{a},f_{b})$ function, with a value of $\beta = 0.25$:
\begin{align*}
\resizebox{0.95\hsize}{!}{%
	$\Theta_{UAF}(f_{a},f_{b})\ \triangleq\ \begin{cases}
	\beta & \parbox[t]{.6\linewidth}{$\text{if } f_{a} \to f_b$ covers more than \hspace*{4em} 2 UAF events in sequence}\\
	1 & \text{otherwise}\end{cases}$
}
\end{align*}

\cref{fig:ex_cg} presents an example of call graph with edges favored
using the above $\Theta_{UAF}$ function.

{\centering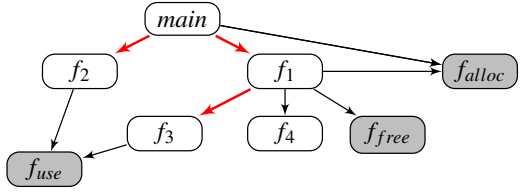
\begin{figure}[t]
	\centering
	\hspace*{-3em}
	\begin{adjustbox}{width=0.4\textwidth}
	\tikzstyle{block} = [font=\footnotesize,rounded corners,rectangle, draw, text width=2em, text centered, inner sep=2pt, minimum height=1.05em]
	\tikzstyle{plain} = [text centered, text width=10em, inner sep=0pt, minimum height=3em]
	\tikzstyle{line} = [draw, -latex',thin]
		
	\begin{tikzpicture}[auto]
	\node[block] (main) [text width=2em] {$main$};
	\node[block] (f2) [below left=0.2cm and 0.3cm of main] {$f_{2}$};
	\node[block] (f1) [below right=0.2cm and 0.3cm of main] {$f_{1}$};
	\node[block] (falloc) [fill=lightgray,node distance=1.5cm,below right=0.2cm and 2.5cm of main] {$f_{alloc}$};
	\node[block] (f3) [below left=0.3cm and 0.5cm of f1] {$f_{3}$};
	\node[block] (f4) [below=0.3cm of f1] {$f_{4}$};
	\node[block] (fuse) [fill=lightgray,below left=0cm and 0.5cm of f3] {$f_{use}$};
	\node[block] (ffree) [fill=lightgray,below right=0.3cm and 0.3cm of f1] {$f_{free}$};

	\path [line,draw=red,thick] (main) -- (f1);
	\path [line,draw=red,thick] (main) -- (f2);
	\path [line] (main) -- (falloc);
	\path [line,draw=red,thick] (f1) -- (f3);
	\path [line] (f1) -- (f4);
	\path [line] (f1) -- (ffree);
	\path [line] (f2) -- (fuse);
	\path [line] (f3) -- (fuse);
	\path [line] (f1) -- (falloc);
	\path [line] (main) -- (falloc);
	\end{tikzpicture}
\end{adjustbox}
\caption{Example of a call graph. Favored edges are in red. 
}\label{fig:ex_cg}
\end{figure}
}\noindent

Finally, we combine our edge weight modification with that of \he{}: \vspace{-0.5em}
\begin{align*}
\resizebox{0.85\hsize}{!}{%
$w_{\uafuzz}(f_{a},f_{b}) \triangleq w_{\he}(f_{a},f_{b}).\Theta_{UAF}(f_{a},f_{b})
$}
\end{align*}

Like \aflgo, we favor the shortest path leading  to the
targets, since it is more likely to involve only a small number of control flow
constraints, making it easier to cover by fuzzing. 
Our distance-based technique therefore considers both calling
relations in general, via $w_{\he}$, and calling relations covering
\uaf\ events in sequence, via $\Theta_{UAF}$.

\subsection{Power Schedule}\label{sec:power}

Coverage-guided fuzzers employ a power schedule (or energy assignment)
to determine the number of extra inputs to generate from a selected
input, which is called the \emph{energy} of the seed. It measures how
long we should spend fuzzing a particular seed. While \afl~\cite{afl}
mainly uses execution trace characteristics such as trace size,
execution speed of the PUT and time added to the fuzzing queue for
seed energy allocation, recent
work~\cite{bohme2016coverage,rawat2017vuzzer,li2019cerebro} including
both directed and coverage-guided fuzzing propose different power
schedules. \aflgo\ employs simulated annealing to assign more energy
for seeds closer to target locations (using the seed distance), while
\he\ accounts for both shorter and longer traces leading to the
targets via a power schedule based on trace distance and similarity at
function level.

We propose here a new power schedule using the intuitions that we
should assign more energy to seeds in these cases:

\noindent\begin{itemize*}
\item seeds that are closer (using the seed distance,
  Section~\ref{sec:usd});\\
\item seeds that are more similar to the target (using the target similarity,
  Section~\ref{sec:tsm});\\
\item {\it seeds that make better decisions at critical code junctions}.  
\end{itemize*}

We define hereafter   a new metric to evaluate the latter case. 

\subsubsection{Cut-edge Coverage Metric}\label{sec:ccm}

To track progress of a seed during the fuzzing process, a fine-grained
approach would consist in instrumenting the execution to compare the
similarity of the execution trace of the current seed with the target
bug trace, at the basic block level. But this method would slow down
the fuzzing process due to high runtime overhead, especially for large
programs.
A more coarse-grained approach, on the other hand, is to measure the
similarity at function level as proposed in
\he~\cite{chen2018hawkeye}. However, a callee can occur multiple times
from different locations of single caller. Also, reaching a target
function does not mean reaching the target basic blocks in this
function.

Thus, we propose the lightweight {\it cut-edge coverage metric}, hitting  a
middle ground between the two aforementioned approaches by 
measuring progress {\it at the edge level}  but  {\it on the critical decision nodes only}. 

{\centering
\makeatletter
\newcommand{\removelatexerror}{\let\@latex@error\@gobble}
\makeatother
\begin{figure}[h]
{\removelatexerror%
\begin{algorithm}[H]\footnotesize
	\caption{Accumulating cut edges}
	\label{alg:acc}
	
	\SetKwInOut{Input}{Input}
	\SetKwInOut{Output}{Output}

	\Input{ Program $P$; dynamic calling tree $T$ of a bug trace}
	\Output{ Set of cut edges $E_{cut}$ }
	\BlankLine
	$E_{cut} \leftarrow \emptyset$;\\
	$nodes \leftarrow$ \textbf{flatten}($T$);\\
	\For{$n \in nodes$ $\wedge$ $pn$ the node before $n$ in $T$}{
		\If{$n.func == pn.func$}{
			$ce$ $\leftarrow$ \textbf{calculate\_cut\_edges}($n.func$, $pn.bb$, $n.bb$);
		} 
		\ElseIf{$pn$ is a call to $n.func$}{ 
			\mbox{$ce$ $\leftarrow$ \textbf{calculate\_cut\_edges}($n.func$, $n.func.entry\_bb$, $n.bb$)};
		}
		$E_{cut} \leftarrow E_{cut} \cup ce$;\\
	}
	\Return $E_{cut}$;
\end{algorithm}%
\vspace{-3mm}\vspace{3pt}      
\begin{algorithm}[H]\footnotesize
	\caption{\textbf{calculate\_cut\_edges} inside a function}
	\label{alg:cutedges}
	
	\SetKwInOut{Input}{Input}
	\SetKwInOut{Output}{Output}
	
	\Input{ A function $f$;\ \ Two basic blocks $bb_{source}$ and $bb_{sink}$ in $f$}
	\Output{ Set of cut edges $ce$ }
	\BlankLine
	$ce \leftarrow \emptyset$; \\
	$cfg \leftarrow$ \textbf{get\_CFG}$(f)$;\\
        $decision\_nodes \leftarrow$ $\{ dn : \exists$ a path $bb_{source} \to^* dn \to^* bb_{sink}$ in $cfg \}$\\
	\For{$dn \in decision\_nodes$}{
		$outgoing\_edges \leftarrow$ \textbf{get\_outgoing\_edges}($cfg$, $dn$);\\
		\For{$edge \in outgoing\_edges$}{
			\If{reachable($cfg$, $edge$, $bb_{sink}$)}{
				$ce \leftarrow ce \cup \{edge\}$;
			}
		}
	}
	\Return $ce$;
      \end{algorithm}}
\end{figure}

}

\begin{definition}{}
  A \emph{cut edge} between two basic blocks \textit{source} and
  \textit{sink} is an outgoing edge of a decision node so that there
  exists a path starting from \textit{source}, going through this edge
  and reaching \textit{sink}. A \emph{non-cut edge} is an edge which
  is not a cut-edge, i.e. for which there is no path from
  \textit{source} to \textit{sink} that go through this edge.
\end{definition}

\cref{alg:acc} shows how cut/non-cut edges are identified in \uafuzz\
given a tested binary program and an expected \uaf\ bug trace. The
main idea is to identify and accumulate the cut edges between all
consecutive nodes in the (flattened) bug trace. 
For instance in the
bug trace of \cref{fig:dct}, we would first compute the cut edges
between $0$ and $1$, then those between $1$ and $2$, etc. As the bug
trace is a sequence of stack traces, most of the locations in the
trace are ``call'' events, and we compute the cut edge from the
function entry point to the call event in that function. However,
because of the flattening, sometimes we have to compute the cut edges
between different points in the same function (e.g. if in the bug
trace the same function is calling \emph{alloc} and \emph{free}, we
would have to compute the edge from the call to \emph{alloc} to the
call to \emph{free}).

\cref{alg:cutedges} describes  how cut-edges are computed inside a
single function. First we have to collect the decision nodes,
i.e. conditional jumps between the source and sink basic blocks. This
can be achieved using a simple data-flow analysis. For each outgoing
edge of the decision node, we check whether they allow to reach the
sink basic block; those that can are cut edges, and the others are
non-cut edges. Note that this program analysis is intra-procedural, so
that we do not need construct an inter-procedural CFG.

Our heuristic is that an input exercising more cut edges and fewer
non-cut edges is more likely to cover more locations of the
target.  

Let $E_{cut}(T)$ be the set of all cut edges of a
program given the expected UAF bug trace $T$. We define the
cut-edge score $e_{s}(s, T)$ of seed $s$ as
\setlength{\abovedisplayskip}{5pt} \setlength{\belowdisplayskip}{5pt}
\begin{equation*}
\resizebox{\hsize}{!}{%
$e_{s}(s, T) \triangleq \mathlarger{\sum_{e\in E_{cut}(T)}\left\lfloor(\log_2hit(e)+1)\right\rfloor - \delta*\!\!\!\sum_{e \notin E_{cut}(T)}\left\lfloor(\log_2hit(e)+1)\right\rfloor}$
}
\end{equation*}
where $hit(e)$ denotes the number of times an edge $e$ is exercised,
and $\delta \in (0,1)$ is the weight penalizing seeds covering non-cut
edges. In our main experiments, we use $\delta = 0.5$ according to our
preliminary experiments. To deal with the path explosion induced by loops,
we use bucketing~\cite{afl}: the hit count is bucketized to small powers of two.

\subsubsection{Energy Assignment}

We propose a power schedule function that assigns energy to a seed
using a combination of the three metrics that we have proposed: the
prefix target similarity metric $t_P(s,T)$ (Section~\ref{sec:tsm}),
the UAF-based seed distance $d(s,T)$ (Section~\ref{sec:usd}), and the
cut-edge coverage metric $e_s(s,T)$ (Section~\ref{sec:ccm}). The idea of our power schedule is to assign energy to a seed $s$
proportionally to the number of targets covered in sequence
$t_{P}(s, T)$, with a corrective factor based on seed distance $d$ and
cut-edge coverage $e_s$. Indeed, our power function (corresponding to
\mbox{\textsc{assign\_energy}} in \cref{alg:dgf}) is defined as:
\begin{equation*}
p(s, T)\quad\triangleq\quad (1 + t_{P}(s, T))\ \times\ \tilde e_{s}(s, T)\ \times\ (1 - \tilde d_{s}(s, T)) 
\end{equation*}

Because their actual value is not as meaningful as the length of the
covered prefix, but they allow to rank the seeds, we apply a min-max
normalization~\cite{aflgo} to get a \emph{normalized seed distance}
($\tilde d_{s}(s, T)$) and \emph{normalized cut-edge score}
($\tilde e_{s}(s, T)$). For example,
$\tilde d_{s}(s, T) = \frac{d_{s}(s,T) - \textit{minD}}{\textit{maxD}
  - \textit{minD}}$ where \textit{minD, maxD} denote the minimum and
maximum value of seed distance so far. Note that both metric scores
are in (0, 1), i.e. can only reduce the assigned energy when their
score is bad.

\subsection{Postprocess and Bug Triage}\label{sec:effic-analysis}
Since \uaf\ bugs are often silent, all seeds generated by a directed fuzzer must {\it a priori} 
be sent to a {\it bug triager} (typically, a profiling tool such as \valgrind) in order to confirm 
 whether they are bug triggering input or not. Yet, this can be extremely expensive as 
fuzzers generate a huge amount of seeds and bug triagers are expensive.

Fortunately, the target similarity metric allows  \uafuzz\  to compute
the sequence of covered targets of each fuzzed input at runtime.  This information is {\it available for free}  
for each  seed once it has been created and executed. We capitalize on it in order to {\it pre-identify} 
 likely-bug triggering seeds, i.e.~seeds that indeed cover the three \uaf\ events in sequence. 
Then, the bug triager is run only over these pre-identified seeds, the other ones being simply discarded -- potentially saving a huge amount of time in bug triaging.
\section{Implementation}
\label{sec:implem}

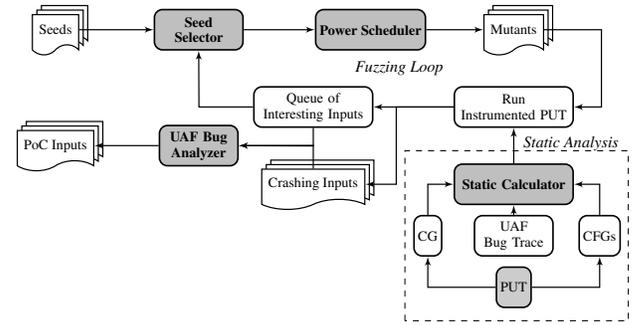
\begin{figure}[htpb]
	\centering
	\begin{adjustbox}{width=\linewidth}
		\tikzstyle{block} = [rectangle, thick, draw, text width=6em, text centered, inner sep=2pt, minimum height=2.5em, font=\scriptsize]
		\tikzstyle{plain} = [text centered, text width=10em, inner sep=0pt, minimum height=3em]
		\tikzstyle{line} = [semithick, draw, -latex']
		\tikzstyle{leg} = [font=\it \tiny]
		\tikzset{
			multidocument/.style={
				shape=tape, semithick,
				font=\scriptsize,text centered,
				draw,
				fill=white,
				tape bend top=none,
				double copy shadow},
			}
		\begin{tikzpicture}[yscale=0.9]
		\draw[draw=black, dashed] (6.1,-5.65) rectangle ++(3.9,3.3);
		\node[text width=, font=\footnotesize] at (9,-2.2) {\textit{Static Analysis}};
		\node[text width=, font=\footnotesize] at (6,-0.75) {\textit{Fuzzing Loop}};
		
		\node[multidocument, text width=, minimum height=2em] (seeds) at (0.0,0) {Seeds};
		\node[block, rounded corners, fill=lightgray, node distance=1.5cm, text width=4em, minimum height=2em] (spi) at (2.5,0) {\textbf{Seed Selector}};
		\node[block, rounded corners, fill=lightgray, node distance=1.5cm, text width=, minimum height=2em] (ps) at (5.5,0) {\textbf{Power Scheduler}};
		\node[multidocument, text width=, minimum height=2em] (mutants) at (8.0,0) {Mutants};
		\node[block, rounded corners, node distance=1.5cm, text width=5.5em, minimum height=2em] (runput) at (8.0,-1.5) {Run\\Instrumented PUT};
		\node[block, rounded corners, node distance=1.5cm, text width=5.5em, minimum height=2em] (queue) at (4.5,-1.5) {Queue of\\Interesting Inputs};
		\node[multidocument, text width=, minimum height=2em] (ci) at (4.5,-3.0) {Crashing Inputs};
		\node[block, rounded corners, fill=lightgray, node distance=1.5cm, text width=3.5em, minimum height=2em] (triage) at (2.5,-2.25) {\textbf{UAF Bug\\Analyzer}};
		\node[multidocument, text width=, minimum height=2em] (poc) at (0.0,-2.25) {PoC Inputs};
		\node[block, rounded corners, fill=lightgray, node distance=1.5cm, text width=5.5em, minimum height=2em] (distance) at (8.0,-3.0) {\textbf{Static Calculator}};
		\node[block, rounded corners, node distance=1.5cm, text width=, minimum height=2em] (cg) at (6.5,-4.0) {CG};
		\node[block, rounded corners, node distance=1.5cm, text width=, minimum height=2em] (cfg) at (9.5,-4.0) {CFGs};
		\node[block, rounded corners, node distance=1.5cm, text width=3.5em, minimum height=2em] (targets) at (8.0,-4.0) {UAF\\Bug Trace};
		\node[block, rounded corners, fill=gray!40, node distance=1.5cm, text width=, minimum height=2em] (put) at (8.0,-5.0) {PUT};
		
		\path [line] (seeds) -- (spi);
		\path [line] (spi) -- (ps);
		\path [line] (ps) -- (mutants);
		\path [line] (runput) -- (queue);
		\path [line] (runput.west) -| ++(-1,0) -- ++(0,-1.5) -- (ci.east);
		\path [line] (triage) -- (poc);
		\path [line] (mutants.east) -| ++(1,-1.5) -- (runput.east);
		\path [line] (queue.west) |- ++(-0.945,0) -- (spi.south);
		\path [line] (queue.south) |- ++(0,-0.35) -- (triage.east);
		\path [line] (ci.north) |- ++(0,0.45) -- (triage.east);
		\path [line] (cg.north) |- (distance.west);
		\path [line] (cfg.north) |- (distance.east);
		\path [line] (put.west) -| (cg.south);
		\path [line] (put.east) -| (cfg.south);
		\path [line] (targets.north) -- (distance.south);
		\path [line] (distance.north) -- (runput.south);
		
		\end{tikzpicture}
	\end{adjustbox}
	\caption{Overview of \uafuzz\ workflow.
	}\label{fig:workflow} 
\end{figure}


We implement our results in a \uaf-oriented binary-level directed fuzzer  named \uafuzz. 
\cref{fig:workflow} depicts an overview of the main components of
\uafuzz. The input of the overall system are a set of initial seeds, the PUT in
binary and target locations extracted from the bug trace. The output is a set of unique bug-triggering inputs. 
The prototype is built upon \afl\ 2.52b~\cite{afl} and QEMU 2.10.0 for fuzzing, and the binary analysis platform \binsec~\cite{binsec} for (lightweight) static analysis. 
These two  components share information such as target locations,
time budget and fuzzing status.  More details in \cref{app:sec:implem}.   

\section{Experimental Evaluation}\label{sec:experiment}
\subsection{Research Questions} \label{sec:rq} 

To evaluate the effectiveness and efficiency of
our approach, we investigate four principal research questions: 

\begin{description}[nosep]
  
\item[RQ1. UAF Bug-reproducing Ability] Can \uafuzz\ outperform other directed 
  fuzzing techniques in terms of \uaf\ bug reproduction in executables? 
\item[RQ2. UAF Overhead] How does \uafuzz\ compare to other directed fuzzing approaches
  w.r.t. instrumentation time and runtime overheads?

\item[RQ3. UAF Triage] How much does \uafuzz\ reduce the number of inputs to be sent to the bug triage step?

\item[RQ4. Individual Contribution] How much does each \uafuzz\ component   
  contribute to the overall results?
\end{description}

\noindent 
We will also evaluate \uafuzz\ in the context of \textit{\textbf{patch testing}}, another important application of directed fuzzing~\cite{bohme2017directed,chen2018hawkeye,peng20191dvul}. 

\subsection{Evaluation Setup} \label{sec:setup} \label{sec:ourbenchmark}

\paragraph{Evaluation Fuzzers} We aim to compare  \uafuzz\ with state-of-the-art directed fuzzers, namely \aflgo~\cite{aflgo} and \he~\cite{chen2018hawkeye}, using \aflqemu\ as a baseline (binary-level coverage-based fuzzing). 
Unfortunately, both \aflgo\ and \he\  work on source code, and while \aflgo\ is open source, \he\ is not available. 
Hence, we {\it implemented binary-level versions} of \aflgo\ and \he, coined as \aflgob\ and  \heb. 
We closely follow the original papers, and, for \aflgo, use the source code as a reference. 
\aflgob\ and \heb\ are implemented on top of \aflqemu, following the
generic architecture of \uafuzz\ but with dedicated distance, seed selection and
power schedule mechanisms.
\cref{tab:fuzzers} summarizes our different fuzzer implementations  and a comparison with their original counterparts. 

{\centering \begin{table}[htpb]
	\centering \footnotesize
	\caption{Overview of main techniques of greybox fuzzers. 
        Our own implementations are marked with  $^\star$.}
	\label{tab:fuzzers}
	\resizebox{\linewidth}{!}{%
		\begin{tabular}{|c|c|c|c|c|c|c|c|}
\hline
Fuzzer &  Directed & Binary? & Distance & Seed Selection & Power Schedule & Mutation \\ \hline\hline

\aflqemu &  \no  & \yes &  -- & \afl & \afl & \afl \\ \hline\hline

\aflgo          &  \yes  & \no  & CFG-based             & $\sim$ \afl               &   Annealing             &   $\sim$ \afl \\ \hline

\aflgob$^\star$ &  \yes  & \yes   &   $\sim$ \aflgo & $\sim$ \aflgo & $\sim$ \aflgo     &   $\sim$ \aflgo  \\ \hline\hline

\he          &  \yes  & \no &   AAFD                  &    distance-based            &   Trace fairness             &    Adaptive \\ \hline

\heb$^\star$  &  \yes  & \yes & $\sim$ \he              &    $\sim$ \he            &    $\approx$ \he            &    $\sim$ \aflgo \\ \hline\hline

\uafuzz$^\star$ &  \yes  & \yes  & UAF-based               &     Targets-based           &    UAF-based         & $\sim$ \aflgo  \\ \hline
\end{tabular} 
}
\end{table}
}

We evaluate the implementation of \aflgob\ (\cref{app:sec:binary-source}, Appendix) and  find it
very close to the original \aflgo\ after accounting for emulation
overhead.

\paragraph{UAF Fuzzing Benchmark} The standard \uaf\ micro benchmark Juliet Test
Suite~\cite{juliet} 
for static analyzers 
is too simple for fuzzing. 
No  macro benchmark actually assesses the
effectiveness of \uaf\ detectors --  the widely used LAVA~\cite{dolan2016lava} 
only contains buffer overflows. Thus, we construct a new \uaf\ benchmark 
according to the following rationale:  

\begin{enumerate}[nosep]
	\item The subjects are real-world  popular and fairly large security-critical programs;  
	
	\item The benchmark includes \uaf\ bugs found by existing fuzzers from the fuzzing
	litterature~\cite{collafl,chen2018hawkeye,bohme2016coverage,afl} or collected
	from NVD\cite{nvd}. Especially, we include \textit{all} \uaf\ bugs found by directed fuzzers; 
	
	\item The bug report provides detailed information (e.g., buggy version and the
	stack trace), so that we can identify target locations for fuzzers. 
\end{enumerate}

\noindent In summary, we have \benchnbug\ known \uaf\ vulnerabilities (2 from  directed fuzzers)
over \benchncode\ real-world C programs whose sizes vary
from 26 Kb to 3.8 Mb. Furthermore, selected programs range  from image
processing to data archiving, video processing and web development. Our
benchmark is therefore representative of different UAF vulnerabilities of
real-world programs. 
\cref{tab:subjects} presents our evaluation benchmark.

{\centering \begin{table}[t]
	\centering \scriptsize
	\caption{Overview of our evaluation benchmark}
	\label{tab:subjects}
	\resizebox{\linewidth}{!}{%
\begin{tabular}{|c|c|c|c|c|c|}
\hline
\multirow{2}{*}{Bug ID} & \multicolumn{2}{c|}{Program} & \multicolumn{2}{c|}{Bug} & {\#Targets} \\ \cline{2-5}
& Project          & Size         & Type      & Crash     &     in trace                    \\ \hline\hline

\texttt{giflib-bug-74}  & GIFLIB                 & 59 Kb                & DF            & \no       & 7      \\ \hline

\texttt{CVE-2018-11496}    & lrzip         & 581 Kb               & UAF           & \no       & 12      \\ \hline

\texttt{yasm-issue-91}    & yasm               & 1.4 Mb               & UAF           & \no       & 19        \\ \hline

\texttt{CVE-2016-4487}  & Binutils          & 3.8 Mb               & UAF           & \yes          & 7       \\ \hline

\texttt{CVE-2018-11416} & jpegoptim        & 62 Kb               & DF            & \no       & 5       \\ \hline

\texttt{mjs-issue-78}     & mjs               & 255 Kb               & UAF           & \no       & 19        \\ \hline

\texttt{mjs-issue-73}   & mjs                 & 254 Kb               & UAF           & \no       & 28        \\ \hline

\texttt{CVE-2018-10685}     & lrzip        & 576 Kb        & UAF           & \no       & 7       \\ \hline

\texttt{CVE-2019-6455}    & Recutils       & 604 Kb               & DF           & \no       & 15       \\ \hline

\texttt{CVE-2017-10686}   & NASM        & 1.8 Mb             & UAF           & \yes      & 10        \\ \hline	

\texttt{gifsicle-issue-122}    & Gifsicle       & 374 Kb               & DF           & \no      & 11      \\ \hline	

\texttt{CVE-2016-3189}   & bzip2          & 26 Kb              & UAF            & \yes      & 5   \\ \hline

\texttt{CVE-2018-20623}   & Binutils          & 1.0 Mb              & UAF            & \no      & 7   \\ \hline

%
%
%

\end{tabular}
}



\end{table}
}

\paragraph{Evaluation Configurations}
We
follow the recommendations for fuzzing evaluations~\cite{klees2018evaluating} and use the same fuzzing configurations and
hardware resources for all experiments. 
Experiments are conducted $10$
times with a time budget depending on the PUT as shown in
\cref{tab:subjects_full_detailed}. We use as input seed either an empty file or  existing valid
files provided by developers. 
We do not use any token dictionary. 
All  experiments were carried out on an  Intel Xeon CPU E3-1505M v6 @
3.00GHz CPU with 32GB RAM and Ubuntu 16.04 64-bit. 

\subsection{UAF Bug-reproducing Ability (RQ1)}
\label{sec:uaf-bug-finding}

\paragraph{Protocol} We compare the different fuzzers on our \benchnbug\  UAF vulnerabilities  using
\emph{Time-to-Exposure} (TTE), i.e.~the time elapsed until first bug-triggering
input, and {\it number of success runs} in which a fuzzer triggers the
bug.
In case a fuzzer  cannot detect the bug within the time budget, the run's TTE
is set to the time budget. Following existing
work~\cite{bohme2017directed,chen2018hawkeye}, we use the \emph{Vargha-Delaney
  statistic ($\hat A_{12}$)} metric~\cite{vargha2000critique} \footnote{Value
  between 0 and 1, the higher the better.  Values above the conventionally large
  effect size of 0.71 are  considered highly
  relevant~\cite{vargha2000critique}.} to assess the confidence that one tool
outperforms another. 
Code coverage
is not relevant for directed fuzzers.  

\begin{figure}[t]
	\includegraphics[width=\linewidth]{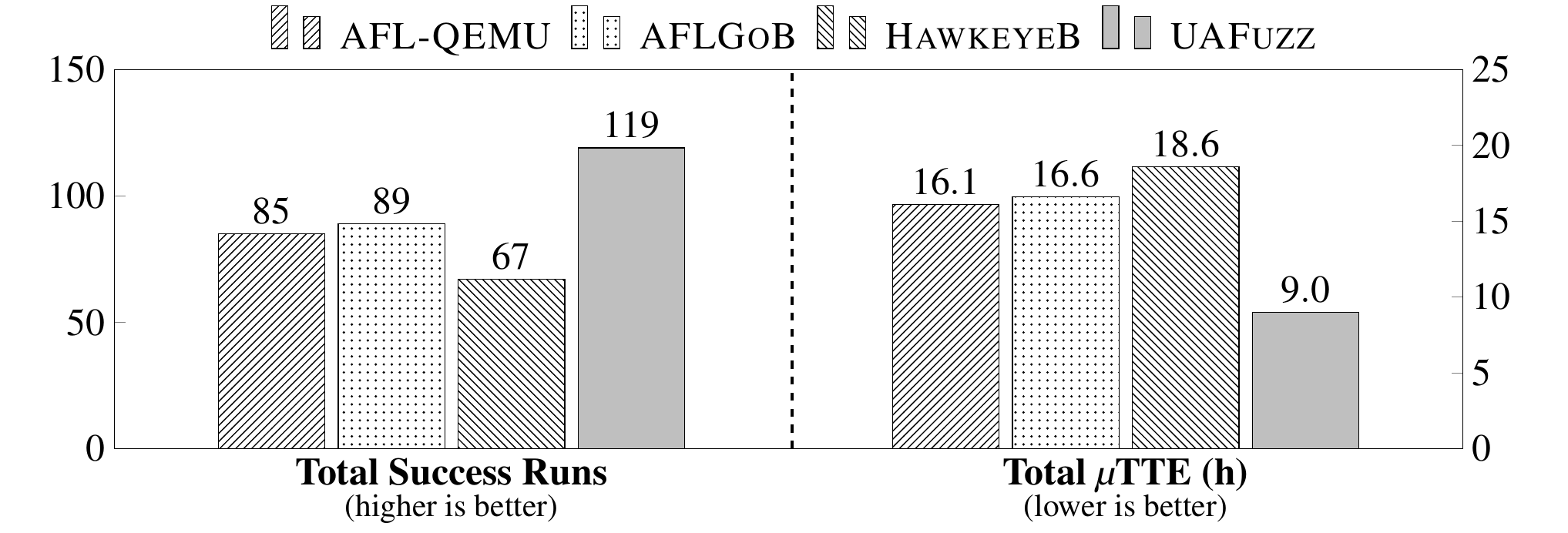}
	\vspace*{-2em}
	\caption{Summary of fuzzing performance (RQ1)}\label{fig:sum}
\end{figure}

\paragraph{Results} \cref{fig:sum} presents a consolidated view of the results (total success runs and TTE -- 
we denote by $\mu$TTE the average TTE observed for each sample over  10 runs).  
\cref{app:sec:rq1} contains additional information: 
detailed statistics per benchmark sample (\cref{tab:ttes}) and consolidated Vargha-Delaney statistics (\cref{tab:a12}).

\cref{fig:sum} (and \cref{tab:ttes,tab:a12}) show that \uafuzz\ clearly outperforms the other fuzzers 
both in total success runs (vs.~2nd best \aflgob: +34\% in total, up to +300\%) and in  TTE (vs.~2nd best \aflgob, total: 2.0$\times$,  avg: 6.7$\times$, max: 43$\times$).
In some
specific cases (see~\cref{tab:ttes}), \uafuzz\ saves roughly 10,000s of TTE over \aflgob\ or 
goes from 0/10 successes to 7/10.     
The $\hat A_{12}$ value of \uafuzz\ against other fuzzers 
is also significantly  above the conventional large effect size 0.71~\cite{vargha2000critique}, as 
shown in \cref{tab:a12}  (vs.~2nd best \aflgob, avg: 0.78, median: 0.80, min: 0.52).

\result{\uafuzz\ \textit{significantly} outperforms state-of-the-art
	directed fuzzers in terms of UAF bugs reproduction with a high
	confidence level.}

Note that performance of \aflgob\ and \heb\ w.r.t.~their original source-level counterparts are  representative (cf.~\cref{app:sec:binary-source}). 

\subsection{UAF Overhead (RQ2)} \label{sec:xp:rq2}

\paragraph{Protocol}  We are interested in both (1) instrumentation-time overhead and (2) runtime overhead. 
For (1), we simply compute the total instrumentation time of \uafuzz\  and we compare it to the instrumentation time of \aflgo.  
For (2), we compute the total number of executions per second  of \uafuzz\  and compare it \aflqemu\ taken as a baseline.  

\begin{figure}[t]
	\includegraphics[width=\linewidth]{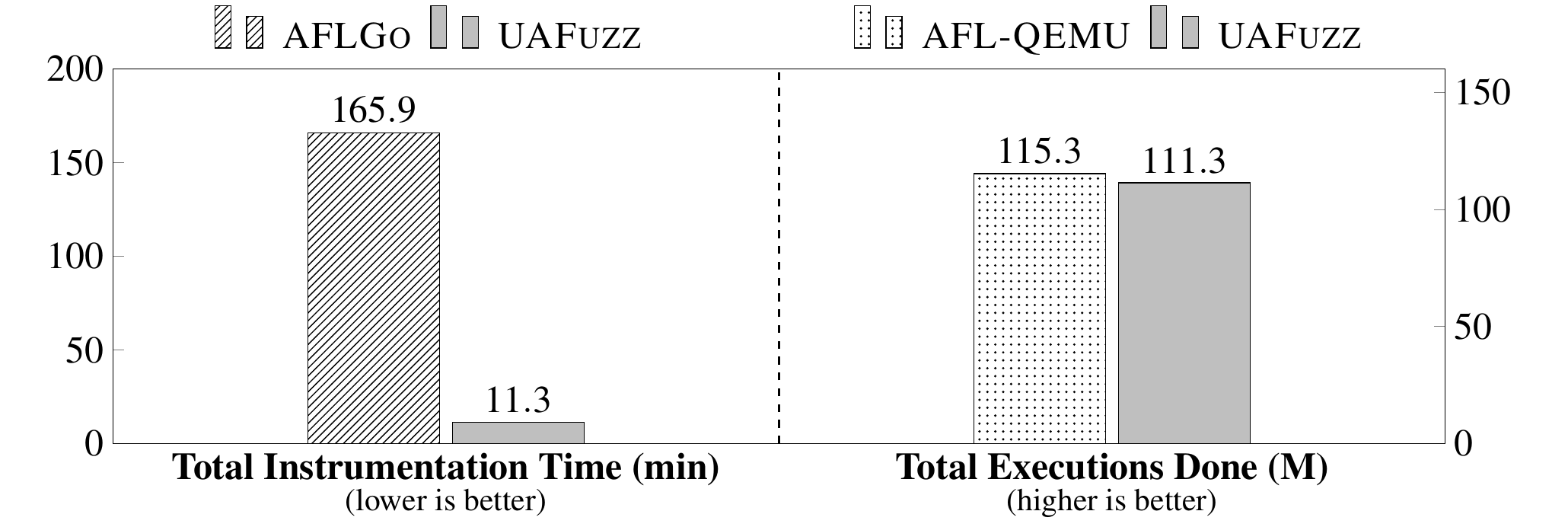}
	\vspace*{-2em}
	\caption{Global overhead (RQ2)}\label{fig:sum_overhead}
\end{figure}

\paragraph{Results}  Consolidated results  for both instrumentation-time and runtime overhead are  presented in \cref{fig:sum_overhead} (number of executions per second is replaced by the total number of executions performed in the same time budget). This figure shows that \uafuzz\ is {\it an order of magnitude faster than the state-of-the-art source-based directed fuzzer \aflgo\ in the instrumentation
phase}, and has almost the same total number of executions per second  as \aflqemu{}.
\cref{app:sec:rq2} contains additional results with detailed instrumentation time  (\cref{fig:instr_time}) and runtime statistics (\cref{fig:total_execs}),  as well as instrumentation time for 
\aflgob\ and \heb\ (\cref{fig:bin_instr_time}).  

\result{\uafuzz\ enjoys both a 
  \textit{lightweight instrumentation time}  and a 
  \textit{minimal runtime overhead}.}

\subsection{UAF Triage (RQ3)}

\paragraph{Protocol} We consider the total number of triaging inputs (number of inputs sent to the triaging step), the triaging inputs rate TIR (ratio between the total number of generated inputs and those sent to triaging) and the total triaging time (time spent within the triaging step). 
Since other
fuzzers cannot identify inputs reaching
targets during the fuzzing process, we conservatively analyze \emph{all} inputs generated by the these fuzzers in the bug triage step (TIR = 1).

\begin{figure}[t]
	\includegraphics[width=\linewidth]{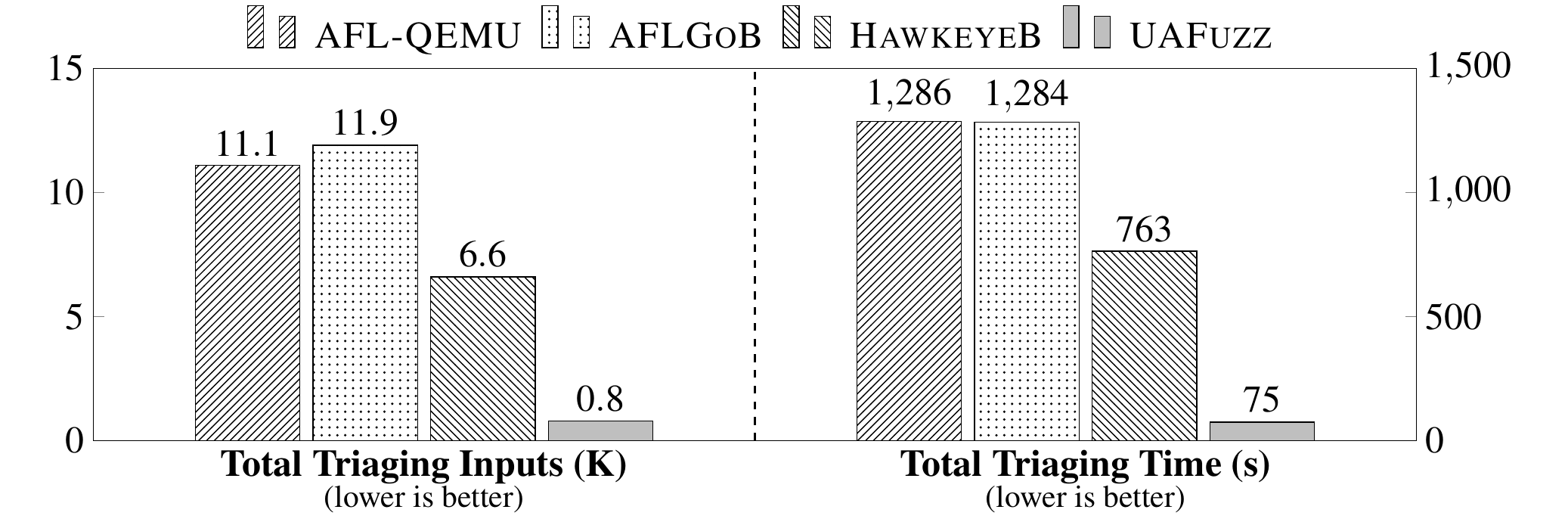}
	\vspace*{-2em}
	\caption{Summary of bugs triage (RQ3)}\label{fig:sum_triage}
\end{figure}

\paragraph{Results} Consolidated results are presented in \cref{fig:sum_triage},  
detailed results in \cref{app:sec:rq3},  
\cref{tab:tir} and  \cref{fig:triage_time}. 

\begin{itemize}[nosep]

\item The TIR of \uafuzz\ is 9.2\%  in total (avg: 7.25\%, median: 3.14\%,  best: 0.24\%, worst: 30.22\%) -- sparing up to 99.76\% of input seeds for confirmation, and is always less than 9\% except for sample \texttt{mjs};

\item \cref{fig:triage_time} shows that \uafuzz\ spends the smallest amount of time in bug triage, i.e. 75s (avg: 6s, min: 1s, max: 24s) for a total speedup of 17$\times$ over \aflgob\ (max: 130$\times$, avg: 39$\times$).   

\end{itemize}

\result{\uafuzz\ reduces a \textit{large} portion (i.e., more than 90\%) of
  triaging inputs in the post-processing phase. Subsequently, \uafuzz\ only
  spends several seconds in this step, winning an order of magnitude compared to standard directed fuzzers.} 

\subsection{Individual Contribution (RQ4)} 

\paragraph{Protocol} We compare  four different versions of our prototype, representing a continuum between \aflgo\ and \uafuzz: 
(1) the basic \aflgo\  represented by \aflgob, 
(2) \aflgobss\ adds our seed selection metric to \aflgob, 
(3) \aflgobds\  adds  the UAF-based function distance to \aflgobss, 
and finally (4) \uafuzz\ adds our dedicated power schedule to \aflgobds.    
%
We consider the previous RQ1 metrics: number of success runs, TTE and Vargha-Delaney.   
Our goal is to assess whether or not these technical improvements do lead to fuzzing performance improvements. 

\begin{figure}[t]
	\includegraphics[width=\linewidth]{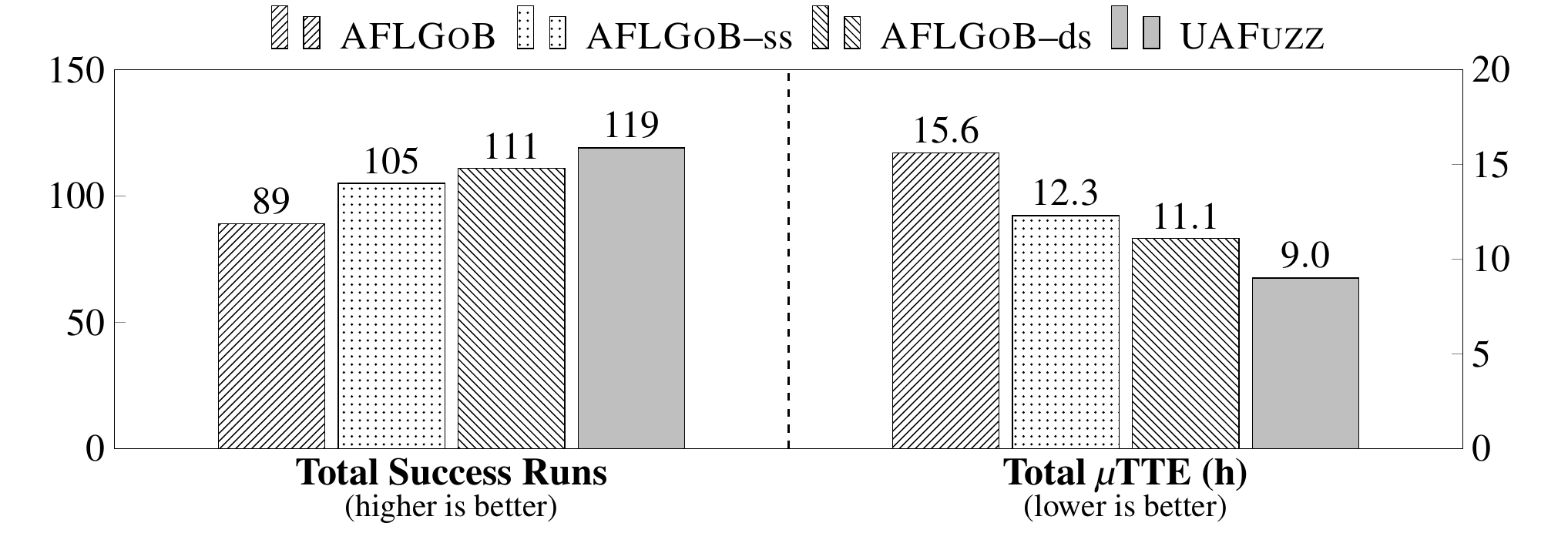}
	\vspace*{-2em}
	\caption{Impact of each components (RQ4)}\label{fig:sum_components}
\end{figure}

\paragraph{Results} Consolidated results for success runs and TTE are represented in \cref{fig:sum_components}. \cref{app:sec:rq4} includes detailed results plus Vargha-Delaney 
metric (\cref{tab:ps}).

 As summarized in~\cref{fig:sum_components}, we can observe that each new component does improve both TTE and number of success runs, leading indeed to fuzzing improvement. 
Detailed results in \cref{tab:ps} with $\hat A_{12}$ values show the same clear trend. 

\result{The UAF-based distance computation, the power scheduling and the seed selection heuristic \textit{individually} contribute to improve  fuzzing performance, and  combining 
them  yield even further improvements, demonstrating their interest and complementarity.} 

{\subsection{Patch Testing \& Zero-days}
\label{sec:zero}
\paragraph{Patch testing}
The idea is to use bug stack traces of {\it known} \uaf\ bugs to guide  testing on the \emph{patched} version of the PUT -- instead of the buggy version as in  bug reproduction. The benefit from  the bug hunting point of view~\cite{gpz} is both to try finding buggy or incomplete patches  
{\it and} to focus testing on {\it a priori} fragile parts of the code, possibly discovering bugs unrelated to the patch itself.  

\paragraph{How to} 
 We follow bug hunting practice~\cite{gpz}. Starting from the recent publicly disclosed \uaf\ bugs of open source programs, we manually identify addresses of relevant call instructions in the reported bug stack traces since the code has been evolved.
We focus mainly on 3  widely-used  programs that have been well fuzzed and maintained by the developers, namely GNU patch,  GPAC and  Perl 5 (737K 
lines of C code and 5 known bug traces in total). We also consider 3 other codes: MuPDF, Boolector  and fontforge (+1,196Kloc).  

\paragraph{Results} 
Overall \uafuzz\ has found and reported \textbf{\nbugs\ new bugs}, including \textbf{\nuafbugs\ new UAF bugs} and \textbf{\ncve\ new CVE} (details in \cref{app:sec:zerodays}, \cref{tab:zeroday}).  {\it At this time, 
\nfixbugs\ bugs    have been fixed by the vendors}. Interestingly, the bugs found in GNU patch (\cref{app:sec:zerodays}) and GPAC were actually {\it buggy patches}. 
\result{\uafuzz\ has been proven effective  in a patch testing setting, allowing to find 
\nbugs\ new bugs (incl.~\ncve\ new CVE) in \ntestedcodes\  widely-used  programs. 
}}\noindent
\subsection{Threats to Validity}\label{sec:discussion}
\paragraph{Implementation} Our prototype is implemented as  part of the binary-level 
code analysis framework 
\binsec{}~\cite{david2016binsec,djoudi2015binsec}, whose efficiency and
robustness have been demonstrated in prior large scale studies on
both adversarial code and managed
code~\cite{bardin_backward-bounded_2017,recoules_ase_2019,david_specification_2016}, and on top of the popular fuzzer
AFL-QEMU. 
Effectiveness and correctness of \uafuzz\ have been assessed on several bug traces from 
real programs, as well as on small samples from the Juliet Test Suite. All 
reported \uaf\ bugs have been manually checked.

\paragraph{Benchmark} 
Our  benchmark is built on both real codes {\it and} real bugs, and encompass several bugs found by recent  fuzzing techniques of well-known open source codes (including all \uaf\ bugs found by directed fuzzers).

\paragraph{Competitors} We consider the best state-of-the-art techniques in directed fuzzing, namely \aflgo~\cite{bohme2017directed} and \he~\cite{chen2018hawkeye}. 
Unfortunately, \he\ is not available and \aflgo\ works on source code only. Thus, we re-implement these technologies in our own framework.
We followed the available information (article, source code if any) as close as possible, and did our best to get precise implementations.
They have both  been checked on real programs and small samples, and the comparison against \aflgo\ source (\cref{app:sec:binary-source}) and  our own \aflgob\ implementation 
is conclusive.

\section{Related Work}\label{sec:rw}
\paragraph{Directed Greybox Fuzzing}
\aflgo~\cite{bohme2017directed} and \he~\cite{chen2018hawkeye} have already been  discussed. 
LOLLY~\cite{liang2019sequence} provides a lightweight instrumentation to measure the sequence basic block coverage of inputs, yet, at the price of a large runtime overhead. 
{\sc SeededFuzz}~\cite{wang2016seededfuzz} seeks to generate a set of initial seeds that improves directed fuzzing performance.
{\sc SemFuzz}~\cite{you2017semfuzz} leverages vulnerability-related texts such as CVE reports to guide  fuzzing.  
{\sc 1dVul}~\cite{peng20191dvul} discovers 1-day vulnerabilities via binary patches. 

{\it \uafuzz\ is the first directed fuzzer tailored to UAF bugs, and one of the very few~\cite{peng20191dvul} able to handle binary code.}

\paragraph{Coverage-based Greybox Fuzzing}
\afl~\cite{afl} is the seminal coverage-guided greybox fuzzer. Substantial efforts have been conducted in the last few years to improve over it \cite{bohme2016coverage,lemieux2018fairfuzz,collafl}.  
Also, many efforts have been fruitfully invested in combining fuzzing with other approaches, such as 
static analysis~\cite{li2017steelix,collafl}, dynamic taint analysis~\cite{rawat2017vuzzer,angora,chen2019matryoshka}, symbolic execution~\cite{stephens2016driller,peng2018t,yun2018qsym} or machine learning~\cite{godefroid2017learn,she2018neuzz}.  

Recently, UAFL~\cite{wang2020typestate} - another independent research effort on the same problem, specialized coverage-guided fuzzing to detect UAFs by finding operation sequences potentially violating a typestate property and then guiding the fuzzing process to trigger property violations. However, this approach relies heavily on the static analysis of source code, therefore is not applicable at binary-level.

{\it Our technique is  orthogonal to all these improvements, they could be reused within \uafuzz\  as is.} 

\paragraph{UAF Detection} Precise static \uaf\ detection is difficult.  \gueb~\cite{gueb} is the only binary-level static analyzer for \uaf.
The technique can be combined with dynamic symbolic execution to generate PoC inputs~\cite{feist2016finding}, yet with scalability issues.  
On the other hand, several \uaf\ source-level static detectors exist, based on  abstract interpretation~\cite{cuoq2012frama}, pointer analysis~\cite{yan2018spatio},   pattern matching~\cite{olesen2014coccinelle}, model checking~\cite{kroening2014cbmc} or demand-driven pointer analysis~\cite{sui2016demand}. 
A common weakness of all static detectors is their inability to infer triggering input -- they rather prove their absence.  
Dynamic \uaf\ detectors mainly rely on  heavyweight instrumentation \cite{caballero2012undangle,nethercote2007valgrind,drmemory} 
and  result in high runtime overhead, even more for closed source programs. 
ASan~\cite{serebryany2012addresssanitizer} performs  lightweight instrumentation, but at source level only.

\paragraph{UAF Fuzzing Benchmark} 
While the Juliet Test Suite~\cite{juliet} (CWE-415, CWE-416)\footnote{Juliet is mostly used for the evaluation of C/C++ static analysis tools.}  contains only too small programs, 
popular fuzzing benchmarks \cite{dolan2016lava,roy2018bug,rode0day,gft,cgc} comprise only very few  \uaf\ bugs. 
Moreover, many of these benchmarks contain either artificial bugs \cite{dolan2016lava,roy2018bug,rode0day,cgc} or artificial programs \cite{juliet}.  

{\it Merging our evaluation benchmark (known UAF) and our new UAF bugs,  we provide the largest fuzzing benchmark dedicated to UAF --   \fullbenchncode\ real codes and  
\fullbenchnbug\  real bugs (\cref{app:sec:benchmark}).   
} 
\section{Conclusion}\label{sec:conclusion} 

\uafuzz\ is the {\it first directed} greybox fuzzing approach tailored to detecting \uaf\ vulnerabilities (in binary) given only the bug stack trace. 
  \uafuzz\ outperforms existing directed fuzzers, 
both in terms of time to bug exposure and 
number of successful runs.
\uafuzz\ has been proven effective in both bug reproduction and patch testing. 

We release the source code of \uafuzz\ and the \uaf\ fuzzing benchmark at:
\begin{center}
	\url{https://github.com/strongcourage/uafuzz}\\
	\url{https://github.com/strongcourage/uafbench}
\end{center}

\section*{Acknowledgement} 
This work was supported by the H2020 project C4IIoT under the Grant Agreement No 833828 and FUI CAESAR.

{\footnotesize
\bibliographystyle{plain}
\bibliography{uafuzz}{}
}
\appendix
\section{UAF Fuzzing Benchmark} \label{app:sec:benchmark}
\cref{tab:subjects_full_detailed} provides additional details about our evaluation benchmark, including program executables under test, buggy commits and fuzzing configurations (test driver, seeds and timeout). 
\cref{tab:subjects_full} presents our full UAF-oriented fuzzing benchmark, incorporating both our evaluation benchmark (bug reproduction) and all CVEs and new bugs used and found during patch testing. 
The benchmark is available~\cite{uafuzzbench}.    
Finally, 
\cref{tab:bench} compares our UAF Fuzzing benchmarks to existing fuzzing benchmarks.

{\centering \begin{table*}[t]
	\centering \scriptsize
	\caption{Detailed view of our evaluation  benchmark}
	\label{tab:subjects_full_detailed}
	\resizebox{\textwidth}{!}{%
\begin{tabular}{|c|c|c|c|c|c|c|c|c|c|c|}
	\hline
	\multirow{2}{*}{Bug ID} & \multicolumn{3}{c|}{Program} & \multicolumn{3}{c|}{Bug} & \multicolumn{4}{c|}{Fuzzing Configuration}
	\\ \cline{2-11}
	& Project & Size                 & Commit                 & Type     & Crash     & Found by     & Test driver                        & Seeds       & Timeout & \# Targets\\ \hline\hline
	
\texttt{giflib-bug-74}  & GIFLIB                 & 59 Kb                & 72e31ff                & DF            & \no       & --            & gifsponge \textless @@                  &``GIF"       &30m   & 7      \\ \hline
	
\texttt{CVE-2018-11496}    & lrzip         & 581 Kb               & ed51e14                & UAF           & \no       & --            & lrzip -t @@                             & lrz files  &15m  & 12      \\ \hline
	
\texttt{yasm-issue-91}    & yasm               & 1.4 Mb               & 6caf151                & UAF           & \no       & AFL          & yasm @@                                 & asm files &1h & 19        \\ \hline

\texttt{CVE-2016-4487}  & Binutils          & 3.8 Mb               & 2c49145                & UAF           & \yes          & AFLFast      & cxxfilt \textless @@                    & empty file  &1h  & 7       \\ \hline

\texttt{CVE-2018-11416} & jpegoptim        & 62 Kb                & d23abf2                & DF            & \no       & --            & jpegoptim @@                            & jpeg files &30m & 5       \\ \hline

\texttt{mjs-issue-78}     & mjs               & 255 Kb               & 9eae0e6                & UAF           & \no       & Hawkeye      & mjs -f @@                               & js files &3h & 19        \\ \hline

\texttt{mjs-issue-73}   & mjs                 & 254 Kb               & e4ea33a                & UAF           & \no       & Hawkeye      & mjs -f @@                               & js files &3h & 28        \\ \hline

\texttt{CVE-2018-10685}     & lrzip        & 576 Kb               & 9de7ccb                & UAF           & \no       & AFL          & lrzip -t @@                             & lrz files  &15m  & 7       \\ \hline
	
\texttt{CVE-2019-6455}    & Recutils       & 604 Kb               & 97d20cc                & DF           & \no       & --          & rec2csv @@                             & empty file  &30m  & 15       \\ \hline

\texttt{CVE-2017-10686}   & NASM        & 1.8 Mb               & 7a81ead                & UAF           & \yes       & CollAFL          & nasm -f bin @@ -o /dev/null      & asm files  & 6h & 10        \\ \hline	

\texttt{gifsicle-issue-122}    & Gifsicle       & 374 Kb               & fad477c                & DF           & \no       & Eclipser          & gifsicle @@ test.gif -o /dev/null      & ``GIF"  & 4h  & 11      \\ \hline	

\texttt{CVE-2016-3189}   & bzip2          & 26 Kb               &  962d606               & UAF            & \yes       & --         & bzip2recover @@ & bz2 files  & 30m  & 5   \\ \hline

\texttt{CVE-2018-20623}   & Binutils          & 1.0 Mb               & 923c6a7                & UAF            & \no       & AFL         & readelf -a @@ & binary files  &1h   & 7      \\ \hline

\end{tabular}
}
\end{table*}
}

{\centering \begin{table}[t]
	\centering \scriptsize
	\caption{Full UAF fuzzing  benchmark}
	\label{tab:subjects_full}
	\resizebox{\linewidth}{!}{%
\begin{tabular}{|c|c|c|c|c|}
	\hline
	\multirow{2}{*}{Bug ID} & \multicolumn{2}{c|}{Program} & \multicolumn{2}{c|}{Bug} \\ \cline{2-5}
	& Project          & Size         & Type      & Crash     \\ \hline\hline
	
	\texttt{giflib-bug-74}  & GIFLIB                 & 59 Kb                & DF            & \no       \\ \hline
	
	\texttt{CVE-2018-11496}    & lrzip         & 581 Kb               & UAF           & \no      \\ \hline
	
	\texttt{yasm-issue-91}    & yasm               & 1.4 Mb               & UAF           & \no      \\ \hline
	
	\texttt{CVE-2016-4487}  & Binutils          & 3.8 Mb               & UAF           & \yes          \\ \hline
	
	\texttt{CVE-2018-11416} & jpegoptim        & 62 Kb               & DF            & \no       \\ \hline
	
	\texttt{mjs-issue-78}     & mjs               & 255 Kb               & UAF           & \no       \\ \hline
	
	\texttt{mjs-issue-73}   & mjs                 & 254 Kb               & UAF           & \no       \\ \hline
	
	\texttt{CVE-2018-10685}     & lrzip        & 576 Kb        & UAF           & \no       \\ \hline
	
	\texttt{CVE-2019-6455}    & Recutils       & 604 Kb               & DF           & \no       \\ \hline
	
	\texttt{CVE-2017-10686}   & NASM        & 1.8 Mb             & UAF           & \yes       \\ \hline	
	
	\texttt{gifsicle-issue-122}    & Gifsicle       & 374 Kb               & DF           & \no       \\ \hline	
	
	\texttt{CVE-2016-3189}   & bzip2          & 26 Kb              & UAF            & \yes       \\ \hline
	
	\texttt{CVE-2016-20623}   & Binutils          & 1.0 Mb              & UAF            & \no      \\ \hline

%
%
%
\hline\hline


\texttt{CVE-2018-6952}   & GNU patch       & 406 Kb         & DF            & \no   \\ \hline
\texttt{CVE-2019-20633}   & GNU patch   & 406 Kb             & DF            & \no     \\ \hline

\texttt{gpac-issue-1340}   & GPAC  & 19.0 Mb         & UAF            & \yes     \\ \hline
\texttt{CVE-2019-20628}   & GPAC   & 14.7 Mb        & UAF            & \yes     \\ \hline
\texttt{CVE-2020-11558}   & GPAC   & 19.0 Mb        & UAF            & \yes     \\ \hline
\texttt{gpac-issue-1440b}   & GPAC  & 19.0 Mb          & UAF            & \yes     \\ \hline
\texttt{gpac-issue-1440c}   & GPAC   & 19.0 Mb          & UAF            & \yes     \\ \hline
\texttt{gpac-issue-1427}   & GPAC    & 19.0 Mb         & UAF            & \yes     \\ \hline

\texttt{perl-1}   & Perl 5       & 8.7 Mb          & UAF        & \yes     \\ \hline
\texttt{perl-2}   & Perl 5       & 8.7 Mb          & UAF       & \yes     \\ \hline
\texttt{perl-3}   & Perl 5       & 8.7 Mb          & UAF     & \yes     \\ \hline
\texttt{perl-4}   & Perl 5       & 8.7 Mb          & UAF            & \yes     \\ \hline

\texttt{MuPDF-issue-701294}   & MuPDF       & 37.6 Mb          & UAF            & \yes     \\ \hline
\texttt{MuPDF-issue-702253}   & MuPDF     & 43.3 Mb           & UAF            & \yes     \\ \hline

\texttt{fontforge-issue-4084}   & fontforge      & 12.4 Mb           & UAF          & \yes     \\ \hline
\texttt{fontforge-issue-4266}   & fontforge & 12.6 Mb                & UAF            & \yes     \\ \hline

\texttt{boolector-issue-41}   & boolector    & 6.9 Mb             & UAF            & \yes     \\ \hline

\end{tabular}
}

\end{table}
}

\section{Implementation} \label{app:sec:implem}
We provide here additional  details about the implementation: 

\begin{itemize}
\item We have implemented a \binsec\ plugin computing \emph{statically} distance
  and cut-edge information, consequently used in the instrumentation of \uafuzz\ -- note that call graph and CFG are retrieved from the \ida\ binary database
  (\ida\ version 6.9~\cite{ida}); 

\item On the {\it dynamic} side, we have modified \aflqemu\ to track covered targets,
  dynamically compute seed scores and power functions;

\item In the end, a small script automates bug triaging. 

\end{itemize}

{\centering \begin{table}[htpb]
	\centering \footnotesize
	\caption{Summary of existing benchmarks.}\label{tab:bench}
\vspace*{-1em}
{\centering\it {\color{orange}$\sim$}\scriptsize: DARPA CGC features  crafted codes and bugs, yet they are supposed to be realistic}
	\resizebox{\linewidth}{!}{%
		\begin{tabular}{|c|c|c|c|c|c|}\hline
\multirow{2}{*}{Benchmark} & \multicolumn{2}{c|}{Realistic} & \multirow{2}{*}{\#Programs} & \multirow{2}{*}{\#Bugs} & \multirow{2}{*}{\#UAF}\\ \cline{2-3}
& Program          & Bug         &                       & &                        \\ \hline\hline

Juliet Test Suite~\cite{juliet}&\no&\no&366&366&366\\
LAVA-1~\cite{dolan2016lava}&\yes&\no&1&69&0\\
LAVA-M~\cite{dolan2016lava}&\yes&\no&4&2265&0\\
{\sc Apocalypse}~\cite{roy2018bug}&\yes&\no&30&30&0\\
{\sc Rode0day}~\cite{rode0day,fasano2019rode0day}&\yes&\no&44&2103&0\\
Google Fuzzer TestSuite~\cite{gft}&\yes&\yes&24&46&3\\
FuzzBench~\cite{fuzzbench}&\yes&\yes&23&23&0\\
DARPA CGC~\cite{cgc}&{\color{orange}$\sim$}&{\color{orange}$\sim$}&296&248&10\\ 



\hline\hline 

UAFuzz Benchmark (evaluation)~\cite{uafuzzbench}&\yes&\yes&\textbf{\benchncode}&\textbf{\benchnbug}&\textbf{\benchnbug}\\

UAFuzz Benchmark (full)~\cite{uafuzzbench}&\yes&\yes&\textbf{\fullbenchncode}&\textbf{\fullbenchnbug}&\textbf{\fullbenchnbug}\\

\hline
\end{tabular} 


}
\end{table}
}\noindent

\paragraph{Memory overhead} \uafuzz\ uses 20 additional bytes of shared memory  
at runtime (per execution): like state-of-the-art source-based DGF we use 8
bytes to accumulate the distance values and 8 bytes to record the number of
exercised basic blocks to store the current seed distance, plus 4
extra bytes for the current seed target trace.

\section{UAF Bug-reproducing Ability (RQ1)} \label{app:sec:rq1}
We present in this section additional results regarding  RQ1, including more 
detailed experimental reports and the {\bf indepth discussion of two notable bugs}:  \texttt{yasm-issue-91} and 
\texttt{CVE-2017-10686} of \texttt{nasm}.  

\paragraph{Experimental results} ~\cref{tab:a12} summarizes the fuzzing performance (details in~\cref{tab:ttes}) of 4 binary-based fuzzers against our benchmark by providing the total number of covered paths, the total number of success runs and the max/min/average/median values of \textit{Factor} and $\hat A_{12}$. \cref{tab:aflgo} compares our fuzzer \uafuzz\ with several variants of directed fuzzers \aflgo.

\begin{table}[h]
	\centering \footnotesize
	\caption{Summary of bug reproduction of \uafuzz\ compared to other fuzzers against our fuzzing benchmark. Statistically significant results $\hat A_{12}$ $\geq 0.71$ are marked as bold.}\label{tab:a12}
	\resizebox{\linewidth}{!}{%
		\includegraphics[width=\linewidth]{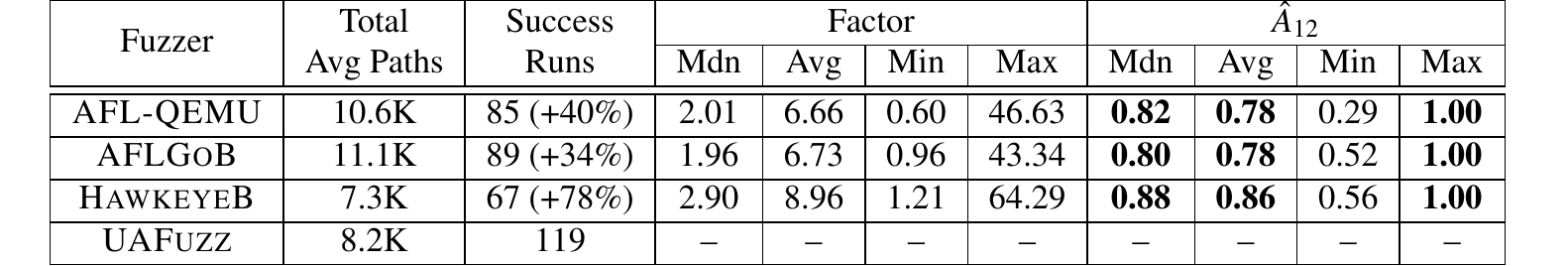}
	}
\end{table}

\begin{table}[h]
	\centering \footnotesize
	\caption{Bug reproduction of \aflgo\ against our benchmark except \texttt{CVE-2017-10686} due to compilation issues of \aflgo. Numbers in red are the best $\mu$TTEs.}
	\label{tab:aflgo}
	\resizebox{\linewidth}{!}{%
		\includegraphics[width=\linewidth]{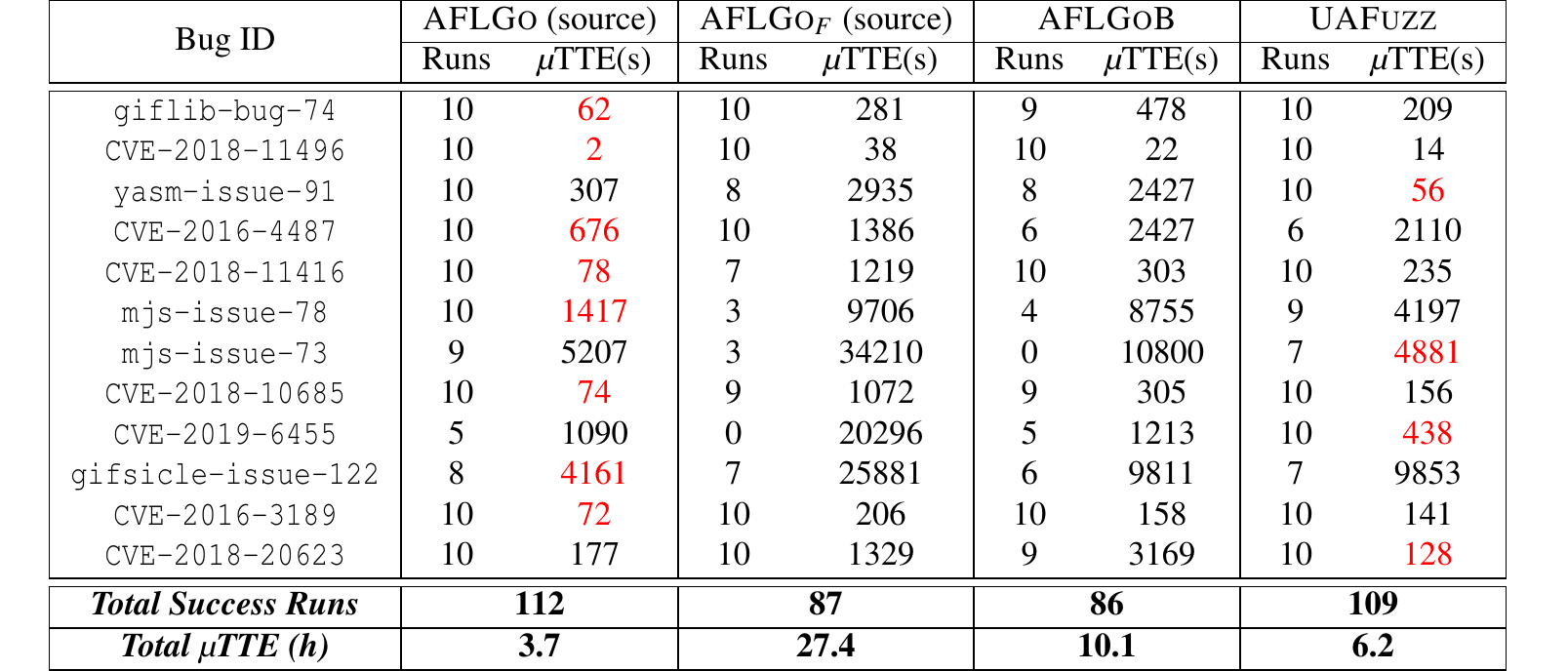}
	}
\end{table}

\begin{table}[h]
	\centering \footnotesize
	\caption{Bug reproduction on $4$ fuzzers against our benchmark. Statistically significant results $\hat A_{12}$ $\geq 0.71$ are marked as bold. \emph{Factor} measures the performance gain as the $\mathbf{\mu}$TTE of other fuzzers divided by the $\mathbf{\mu}$TTE of \uafuzz.}
	\label{tab:ttes}
	\resizebox{\linewidth}{!}{%
		\includegraphics[width=\linewidth]{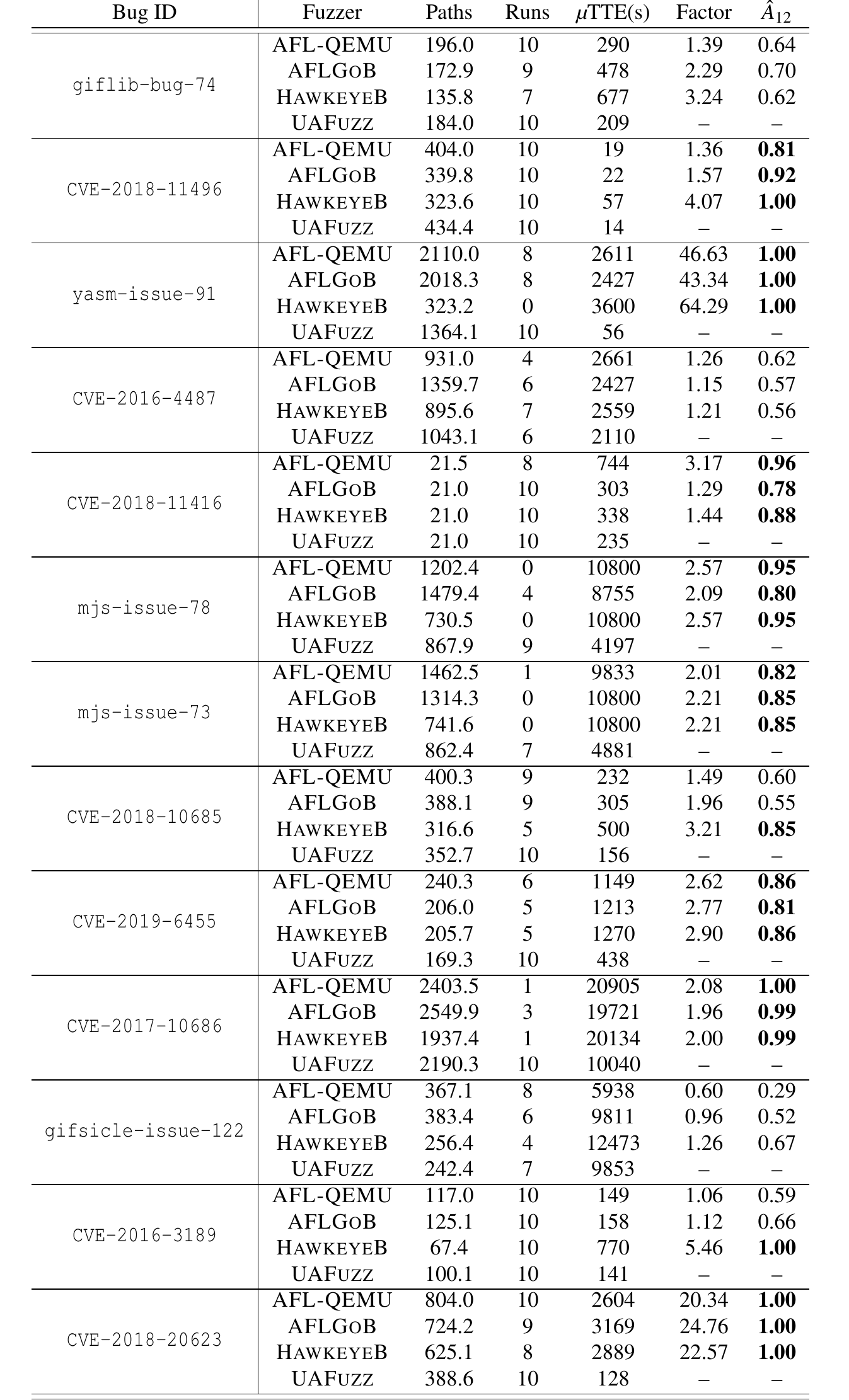}
	}
\end{table}

\begin{figure}[t]
	\includegraphics[width=\linewidth]{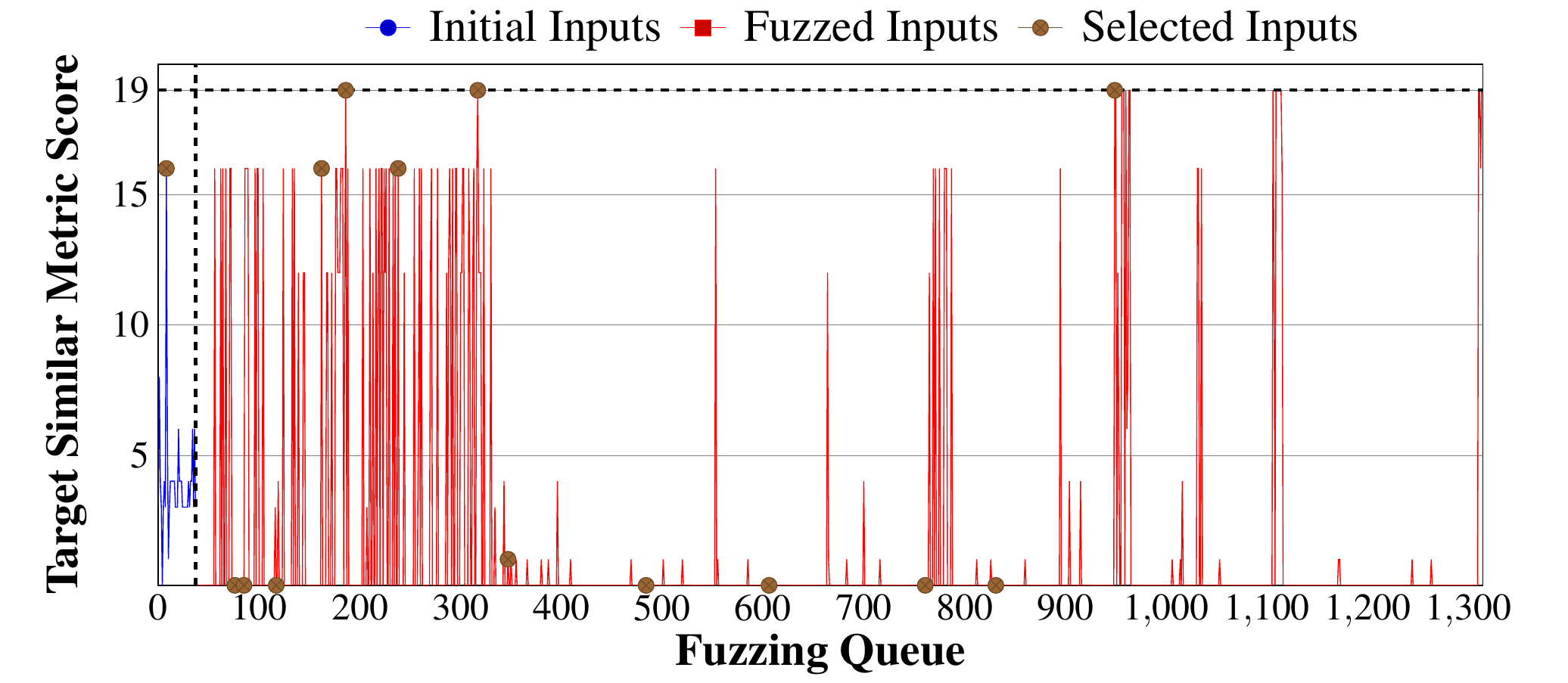}
	\vspace*{-2em}
	\caption{Fuzzing queue of \uafuzz\ for \texttt{yasm-issue-91}. Selected inputs to be mutated are highlighted in brown. Potential inputs are in the horizontal dashed line.}\label{fig:cs_yasm}
\end{figure}

\paragraph{Zoom on two bugs} We discuss the case of \textbf{\texttt{yasm-issue-91}}, where in all 10 runs, \uafuzz\ needs only in a few seconds to reproduce the bug, thus gains a speedup of 43$\times$ over the second best tool \aflgob\ with a high confidence (i.e., $\hat A_{12}$ is 1 against other fuzzers). \cref{fig:cs_yasm} depicts the fuzzing queue of our fuzzer \uafuzz\ for the case study in one run. We can see that our seed selection heuristic first selects the most promising inputs among the set of initial test suite (i.e., the most left circle point). As this input also has the biggest cut-edge score among the initial seeds, \uafuzz\ spends enough long time to mutate this input and thus eventually discovers the first potential input whose execution trace is similar to the expected trace. Then, two first potential inputs covering in sequence all 19 targets are selected to be mutated by \uafuzz\ during fuzzing. Consequently, \uafuzz\ could trigger the bug at the third potential input (i.e., the 954th input in the fuzzing queue). Overall in 10 runs the first bug-triggering input of \uafuzz\ is the 1019th on average, while for \aflqemu\ and \aflgob\ they detect the bug much slower, at the 2026th and 1908th input respectively. The main reason is that other tools spend more time on increasing the code coverage by going through all initial seeds in the fuzzing queue. In particular, as \aflgob\ aims to first explore more paths in the exploration phase, it is more likely that directed fuzzers that are mainly based on the seed distance metric like \aflgob\ skip or select the input after long time. Although both \aflqemu\ and \aflgob\ could find the bug in 8 and 10 runs and discover substantially more paths than our fuzzer, the TTE values of these tools are clearly much more larger than \uafuzz's TTE.

Another interesting case study illustrating the effectiveness of our technique is \textbf{\texttt{CVE-2017-10686}} of the program \texttt{nasm} version 2.14rc0, reported by CollAFL~\cite{collafl}. When assembling the input text file, \texttt{nasm} converts each line into tokens that are allocated in function \texttt{new\_token} and then freed in function \texttt{detoken} to detokenize the line and emit it. However, the buggy version of \texttt{nasm} also frees token's text even when the text has not been modified before, causing the UAF bug. While \uafuzz\ found the UAF bug in all 10 runs, \aflqemu\ and \aflgob\ only produced the bug-triggering input in 1 and 3 runs respectively. Overall \uafuzz\ significantly outperforms other fuzzers and achieves a strong confidence of being faster ($\hat A_{12}$ is 0.99 against \aflgob\ and 1 against the rest). Specifically, \uafuzz\ saves roughly 10,000s to reproduce this bug.

\section{Regarding implementations of \aflgob\ and \heb} \label{app:sec:binary-source}

\paragraph{Comparison between \aflgob\ and  source-based \aflgo} \label{sec:aflgo-aflgob}

\begin{figure}[t]
	\includegraphics[width=\linewidth]{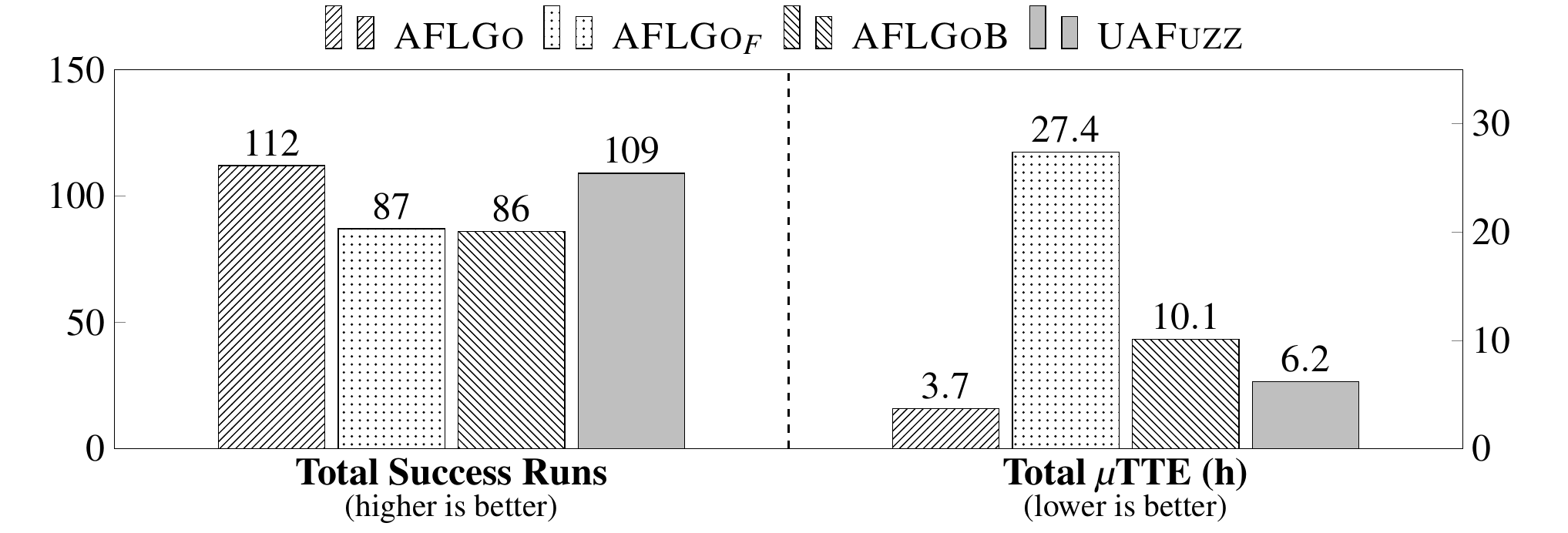}
	\vspace*{-2em}
	\caption{Summary of fuzzing performance of 4 fuzzers against our benchmark, except \texttt{CVE-2017-10686} due to compilation issues of \aflgo.}\label{fig:sum_aflgo}
\end{figure}

We want to evaluate how close our implementation of \aflgob\ is from the original 
 \aflgo, in order to  assess the degree of confidence we can have in our results -- we do not do it for \heb\ as \he\ is not available.  

\aflgo\ unsurprisingly performs better  than \aflgob\ and  \uafuzz\ 
(\cref{fig:sum_aflgo}, \cref{tab:aflgo} in Appendix). This is largely due to the emulation runtime overhead
of QEMU, a well-documented fact.  Still, {\it surprisingly enough}, \uafuzz\ can
 find the bugs faster than \aflgo\ in 4 samples, demonstrating its 
efficiency. 

Yet, more interestingly, \cref{fig:sum_aflgo} also shows that once  emulation overhead \footnote{We
  estimate for each sample an overhead factor $f$ by comparing the number of
  executions per second in both \afl\ and \aflqemu, then multiply the computation
  time of \aflgo\ by $f$ -- $f$ varies from 2.05 to 22.5 in our samples.}  is
taken into account (yielding \aflgof, the expected {\it binary-level} performance of \aflgo), 
then \aflgob\ is in line with  \aflgof\
(and even shows better TTE) --  \uafuzz\ even significantly outperforms
\aflgof.

\begin{table}[h]
	\centering \footnotesize
	\caption{Bug reproduction of \aflgo\ against our benchmark except \texttt{CVE-2017-10686} due to compilation issues of \aflgo. Numbers in red are the best $\mu$TTEs.}
	\label{tab:aflgo}
	\resizebox{\linewidth}{!}{%
		\includegraphics[width=\linewidth]{figures/aflgo.pdf}
	}
\end{table}

\result{Performance of \aflgob\ is in line with the original \aflgo\
  once QEMU overhead is taken into account, allowing a fair comparison with
  \uafuzz. \uafuzz\ nonetheless performs relatively well on \uaf\
  compared with the source-based directed fuzzer \aflgo, demonstrating the
  benefit of our original fuzzing mechanisms. }

\paragraph{About performance of \heb\ in RQ1}  \heb\ performs significantly worse than \aflgob\ and \uafuzz\ in \cref{sec:uaf-bug-finding}. We cannot compare \heb\ with \he\ as \he\ is not available.   
Still, we investigate that issue and found that this is mostly due to a large runtime overhead  
spent calculating the target similarity metric. Indeed, according to the \he\ 
original paper \cite{chen2018hawkeye}, this computation involves some {\it quadratic computation}   
over the {\it total number of functions} in the code under test. On our samples  this number  quickly becomes  important (up to 772) 
 while the number of targets (\uafuzz) remains small (up to 28).  A few examples:  \texttt{CVE-2017-10686}: 772 functions vs 10 targets; \texttt{gifsicle-issue-122}: 516 functions vs 11 targets; \texttt{mjs-issue-78}: 450 functions vs 19 targets.  
Hence, we can conclude that on our samples the  performance of \heb\ are in line with what is 
expected from \he\ algorithm.  

\section{UAF Overhead (RQ2)} \label{app:sec:rq2}
\begin{figure}[h]
	\includegraphics[width=\linewidth]{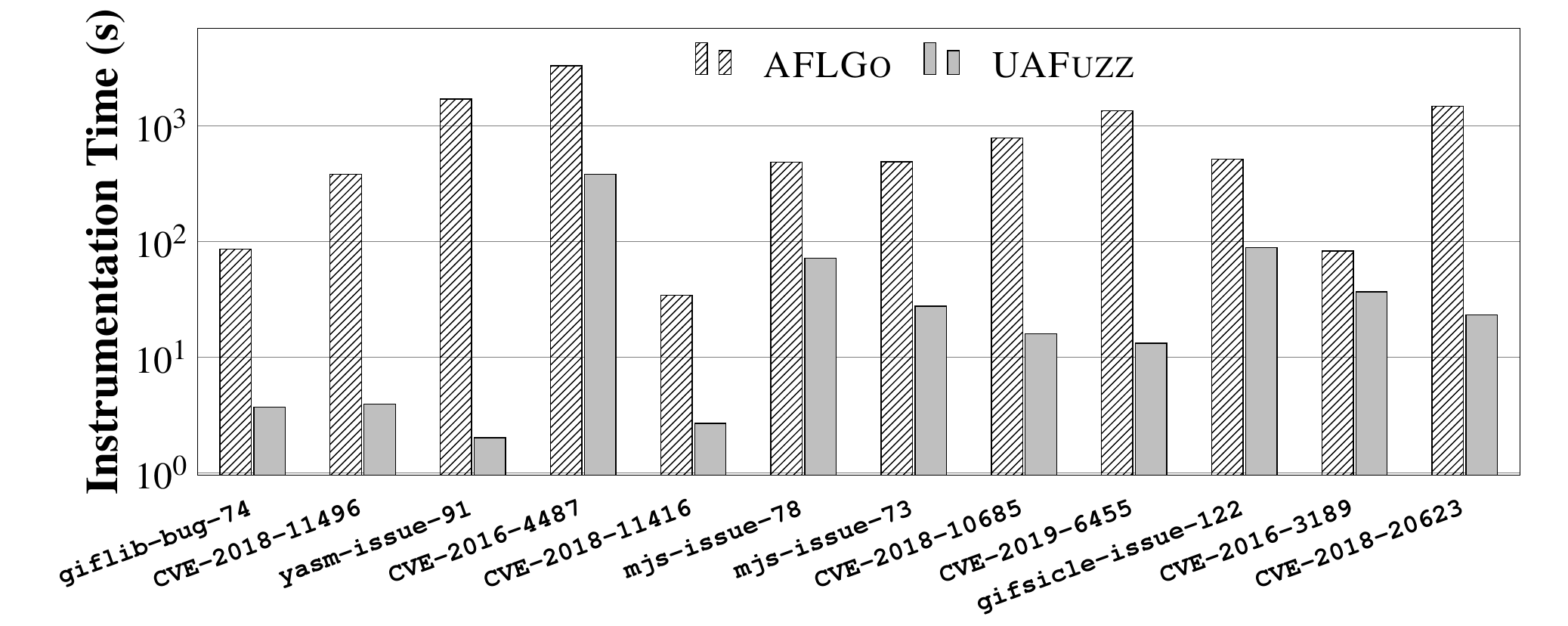}
	\vspace*{-2em}
	\caption{Average instrumentation time in seconds (except \texttt{CVE-2017-10686} due to compilation issues of \aflgo).}\label{fig:instr_time}
\end{figure}

\begin{figure}[h]
	\includegraphics[width=\linewidth]{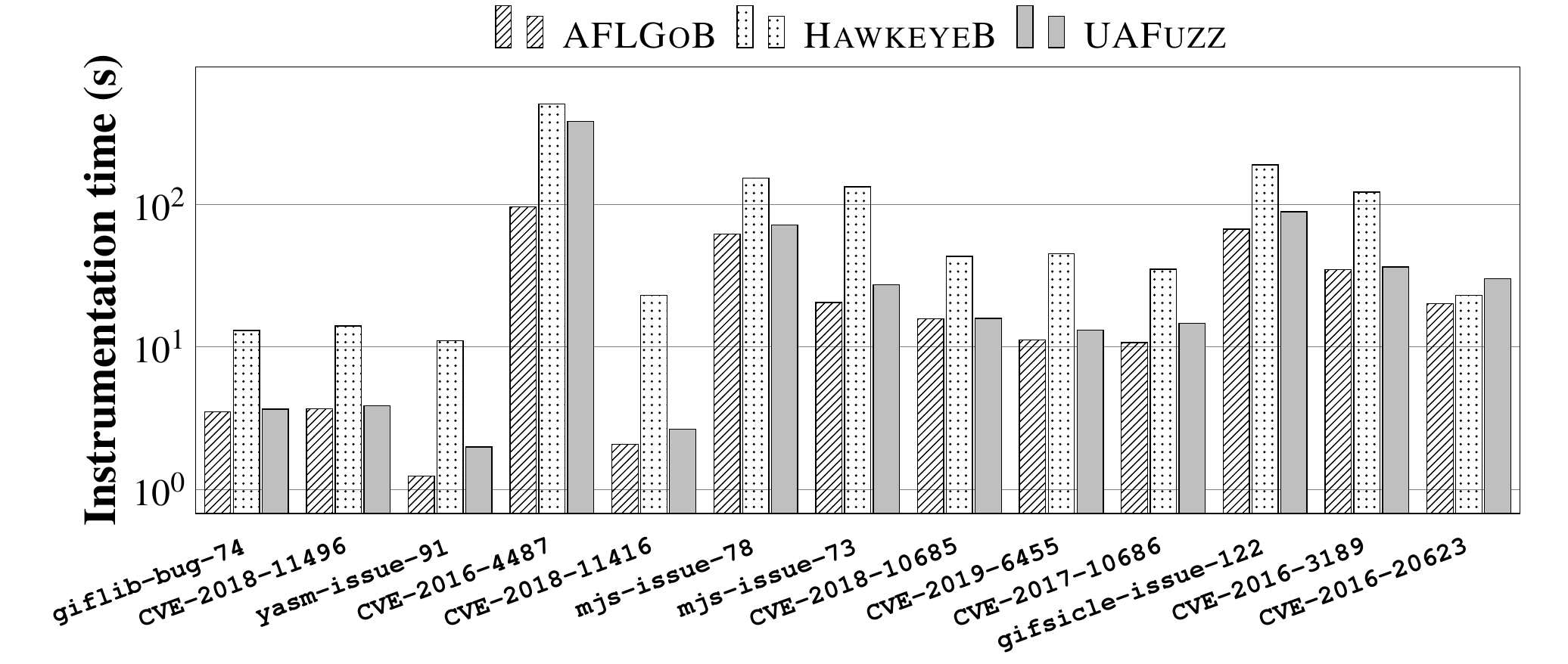}
	\vspace*{-2em}
	\caption{Average instrumentation time in seconds.}\label{fig:bin_instr_time}
\end{figure}

\begin{figure}[h]
	\includegraphics[width=\linewidth]{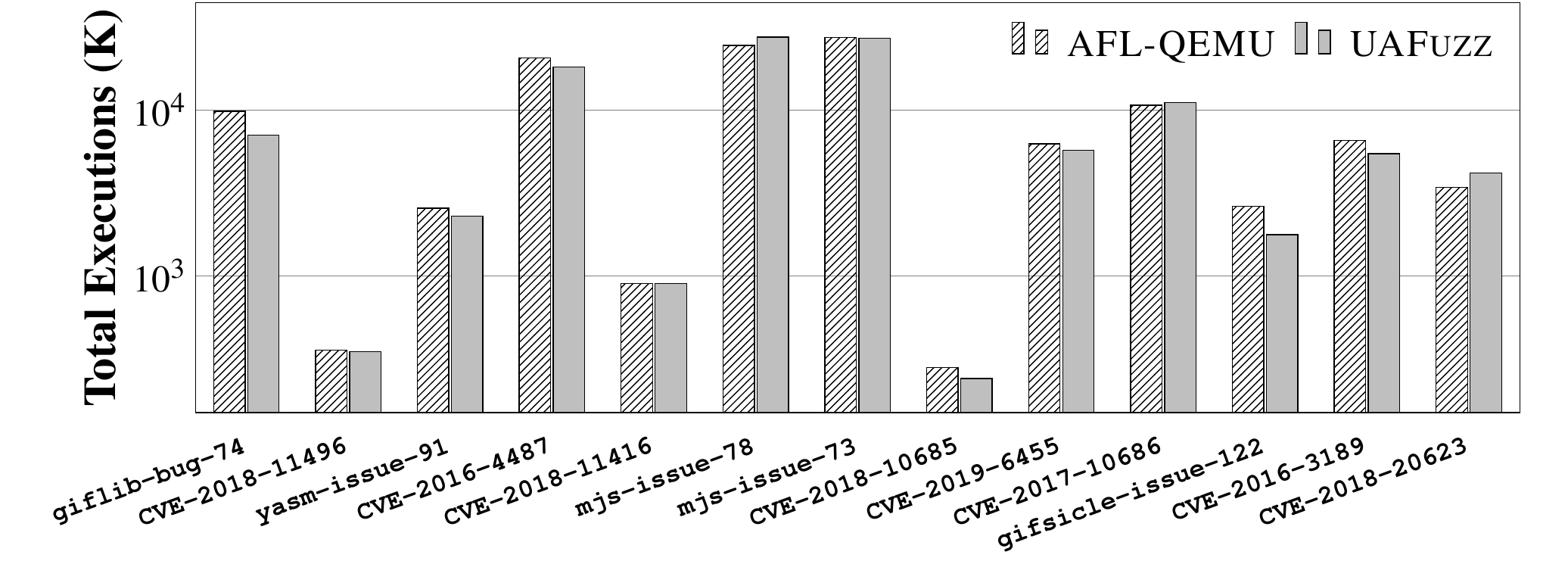}
	\vspace*{-2em}
	\caption{Total executions done in all runs.}\label{fig:total_execs}
\end{figure}

\paragraph{Additional data} We first provide additional results for RQ2. Figures~\ref{fig:instr_time} and \ref{fig:bin_instr_time} compare the average instrumentation time between, respectively, \uafuzz\ and the source-based directed fuzzer \aflgo{}; and \uafuzz{} and the two binary-based directed fuzzers \aflgob\ and \heb. \cref{fig:total_execs} shows the total execution done of \aflqemu\ and \uafuzz\ for each subject in our benchmark.

\paragraph{Detailed results} We now discuss experimental results regarding overhead in more depth than what was done in \cref{sec:xp:rq2}.

\begin{itemize}[nosep]

\item \cref{fig:sum_overhead,fig:instr_time} show that \uafuzz\ is {\it an order of magnitude faster  than the
state-of-the-art source-based directed fuzzer \aflgo\ in the instrumentation
phase} (14.7$\times$ faster in total). For example, \uafuzz\ spends only 23s (i.e., 64$\times$ less than \aflgo) in
processing the large program  \texttt{readelf} of Binutils;
\item \cref{fig:sum_overhead,fig:total_execs} show that \uafuzz\ has almost the same total number of executions per second  as \aflqemu\ 
(-4\% in total, -12\% in average), meaning that its overhead  is  negligible.  

\item \cref{fig:bin_instr_time} shows that \heb\  is sometimes significantly slower than \uafuzz\ (2$\times$).   
This is  mainly because of the cost of target function trace closure calculation on large examples with many functions (cf.~\cref{sec:uaf-bug-finding}). 
\end{itemize}

\section{UAF Triage (RQ3)} \label{app:sec:rq3} 

\begin{figure}[h]
	\includegraphics[width=\linewidth]{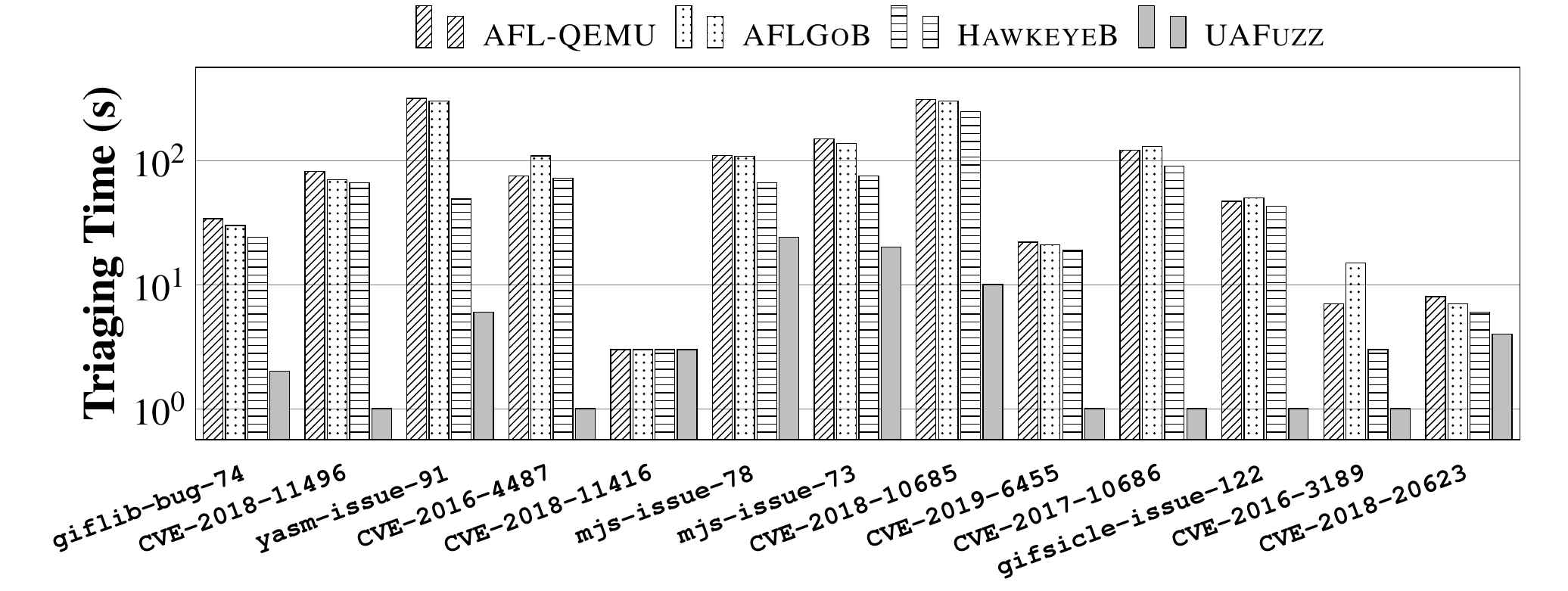}
	\vspace*{-2em}
	\caption{Average triaging time in seconds.}\label{fig:triage_time}
\end{figure}

We provide additional results for RQ3: \cref{fig:triage_time} and \cref{tab:tir} show the average triaging time and number of triaging inputs (including TIR values for \uafuzz) of 4 fuzzers against our benchmark. 

\begin{table}[h]
	\centering \footnotesize
	\caption{Average number of triaging inputs of $4$ fuzzers against our tested subjects. For \uafuzz, the TIR values are in parentheses.}
	\label{tab:tir}
	\resizebox{\linewidth}{!}{%
		\includegraphics[width=\linewidth]{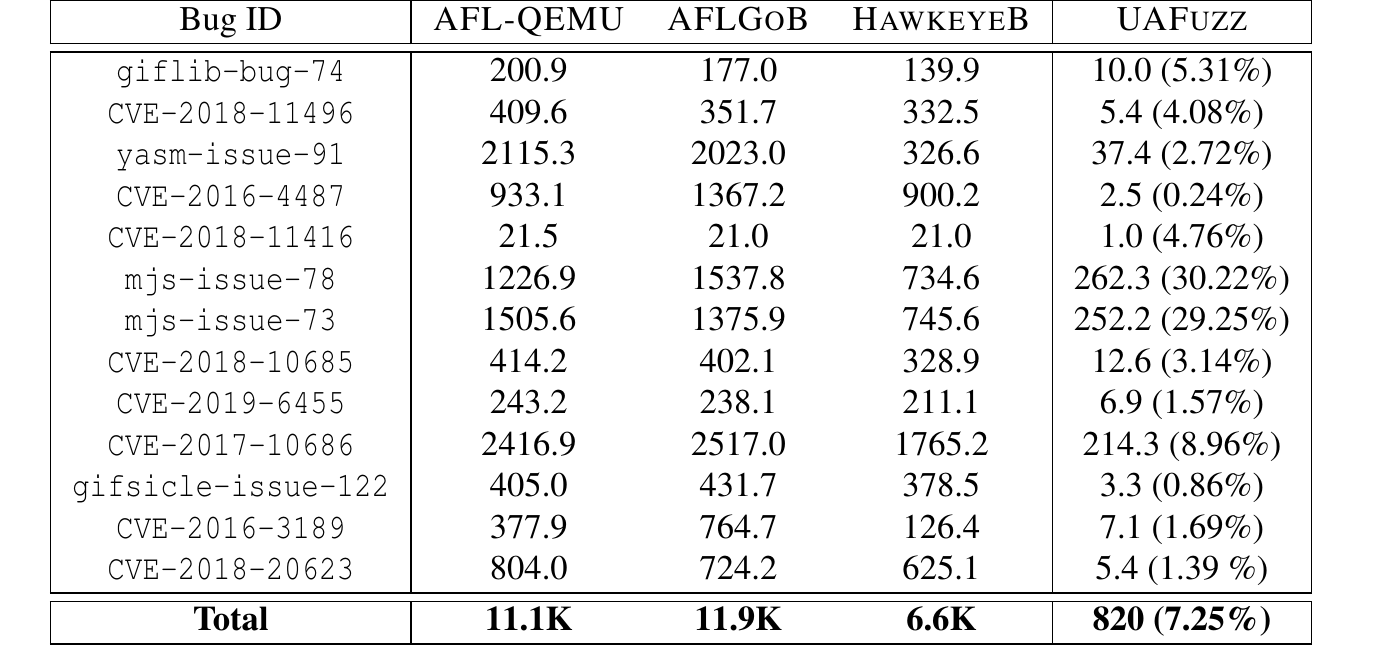}
	}
\end{table}

\section{Individual Contribution (RQ4)} \label{app:sec:rq4}

We provide additional results for RQ4.~\cref{tab:ps} shows the fuzzing performance of 2 \aflgob-based variants \aflgobss\ and \aflgobds\ compared to \aflgob\ and our tool \uafuzz\ against our benchmark.

\begin{table*}[t]
	\centering \footnotesize
	\caption{Bug reproduction on $4$ fuzzers against our benchmark. \aaa\ and \au\ denote the Vargha-Delaney values of \aflgob\ and \uafuzz.  
		Statistically significant results for $\hat A_{12}$ (e.g., \aaa\ $\leq 0.29$ or \au\ $\geq 0.71$) are in bold. Numbers in red are the best $\mu$TTEs.
	}
	\label{tab:ps}
	\resizebox{0.9\linewidth}{!}{%
		\includegraphics[width=\linewidth]{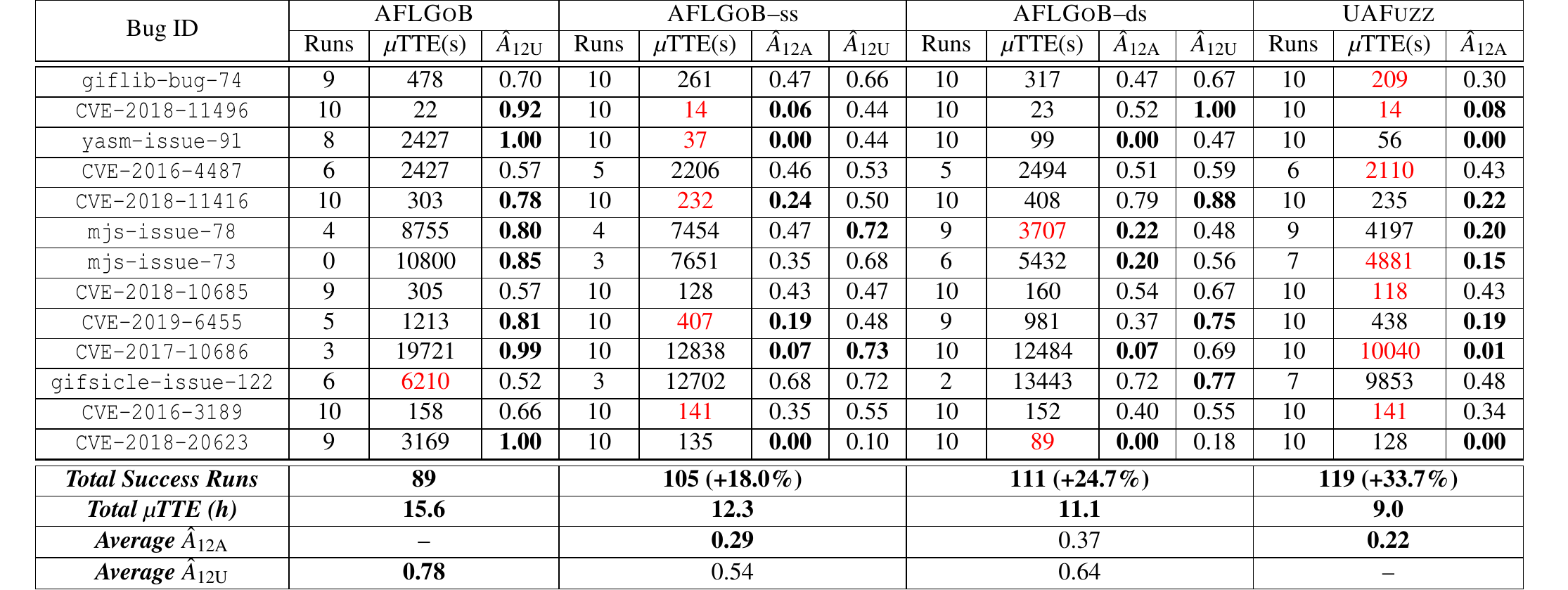}
	}
\end{table*}

\section{Patch Testing \& Zero-days} \label{app:sec:zerodays}

We provide additional results for patch testing (~\cref{tab:zeroday}), as well as a {\bf detailed discussion on the GNU Patch buggy patch}.  

\paragraph{Zoom: GNU Patch buggy patch} We use \texttt{CVE-2018-6952}~\cite{cvedfpatch}  to demonstrate the effectiveness of \uafuzz\ in exposing unknown UAF vulnerabilities. 
GNU patch~\cite{gnupatch} takes a patch file containing a list of differences and applies them to the original file. \cref{lst:patchsrc} shows the code fragment of \texttt{CVE-2018-6952} which is a double free in the latest version 2.7.6 of GNU patch. Interestingly, by using the stack trace of this CVE as shown in~\cref{fig:CVE-2018-6952}, \uafuzz\ successfully discovered an \emph{incomplete bug fix}~\cite{dfpatch} in the latest commit \texttt{76e7758}, with a slight difference of the bug stack trace (i.e., the call of \texttt{savebuf()} in \texttt{another\_hunk()}).

{\centering \begin{lstlisting}[float,floatplacement=tb,language=customC,caption={Code fragment of GNU patch pertaining to the UAF vulnerability \texttt{CVE-2018-6952}.},captionpos=b,label={lst:patchsrc}]
@\textcolor{olive}{\bfseries File: src/patch.c}@
int main (int argc, char **argv) {...
	while (0 < (got_hunk = another_hunk (diff_type, reverse))) { /* Apply each hunk of patch */ ... }@\label{cs:another_hunk}@
...}

@\textcolor{olive}{\bfseries File: src/pch.c}@
int another_hunk (enum diff difftype, bool rev) { ...
	while (p_end >= 0) {
		if (p_end == p_efake) p_end = p_bfake;
		else free(p_line[p_end]); /* Free and Use event */@\label{cs:free}@
		p_end--;
	} ...
	while (p_end < p_max) { ...
		switch(*buf) { ...@\label{cs:switch}@	
		case '+': case '!': /* Our reported bug */ ...
			p_line[p_end] = savebuf (s, chars_read);@\label{cs:our_alloc}@ ...
		case ' ': /* CVE-2018-6952 */ ...
			p_line[p_end] = savebuf (s, chars_read); ...
		...}
	...}
... }

@\textcolor{olive}{\bfseries File: src/util.c}@
/* Allocate a unique area for a string. */
char *savebuf (char const *s, size_t size) { ...
	rv = malloc (size); /* Alloc event */ ...@\label{cs:malloc}@
	memcpy (rv, s, size);
	return rv;
}
\end{lstlisting}

}

\begin{figure}[ht!]\ttfamily\scriptsize
	\begin{tabular}{@{}l@{}}
		==330== Invalid free() / delete / delete[] / realloc()\\
		==330==    at 0x402D358: free (in vgpreload\_memcheck-x86-linux.so)\\
		==330==    by 0x8052E11: another\_hunk (pch.c:1185)\\
		==330==    by 0x804C06C: main (patch.c:396)\\
		==330==  Address 0x4283540 is 0 bytes inside a block of size 2 free'd\\
		==330==    at 0x402D358: free (in vgpreload\_memcheck-x86-linux.so)\\
		==330==    by 0x8052E11: another\_hunk (pch.c:1185)\\
		==330==    by 0x804C06C: main (patch.c:396)\\
		==330==  Block was alloc'd at\\
		==330==    at 0x402C17C: malloc (in vgpreload\_memcheck-x86-linux.so)\\
		==330==    by 0x805A821: savebuf (util.c:861)\\
		==330==    by 0x805423C: 
		\color{red!80}{\textbf{another\_hunk (pch.c:1504)}}\\
		==330==    by 0x804C06C: main (patch.c:396)\\
	\end{tabular}
	\caption{The bug trace of CVE-2018-6952 produced by \valgrind.}\label{fig:CVE-2018-6952}
\end{figure}

Technically, GNU patch takes an input patch file containing multiple hunks (line~\ref{cs:another_hunk}) that are split into multiple strings using special characters as delimiter via \texttt{*buf} in the switch case (line~\ref{cs:switch}). GNU patch then reads and parses each string stored in \texttt{p\_line} that is dynamically allocated on the memory using \texttt{malloc()} in \texttt{savebuf()} (line~\ref{cs:malloc}) until the last line of this hunk has been processed. Otherwise, GNU patch deallocates the most recently processed string using \texttt{free()} (line~\ref{cs:free}). Our reported bug and \texttt{CVE-2018-6952} share the same \emph{free} and \emph{use} event, but have a different stack trace leading to the same \emph{alloc} event. Actually, while the PoC input generated by \uafuzz\ contains two characters `\texttt{!}', the PoC of \texttt{CVE-2018-6952} does not contain this character, consequently the case in line~\ref{cs:our_alloc} was previously uncovered, and thus this CVE had been incompletely fixed. 
This case study shows the importance of producing different unique bug-triggering inputs to favor the repair process and help complete bug fixing.

{\centering \begin{table*}[htpb]
	\centering
	\caption{Summary of zero-day vulnerabilities reported by our fuzzer.}
	\label{tab:zeroday}
\begin{adjustbox}{width=0.9\textwidth,center}
		\begin{tabular}{|c|c|c|c|c|c|c|c|c|}
\hline
Program & Code Size & Version (Commit) & Bug ID & Vulnerability Type & Crash & Vulnerable Function & Status & CVE \\ \hline\hline

\multirow{15}{*}{GPAC} & \multirow{15}{*}{545K}
&0.7.1 (987169b)&\#1269&\textbf{User after free}&\no&gf\_m2ts\_process\_pmt&Fixed&CVE-2019-20628\\\cline{3-9}

&&0.8.0 (56eaea8)&\#1440-1&\textbf{User after free}&\no&gf\_isom\_box\_del&Fixed&CVE-2020-11558\\\cline{3-9}

&&0.8.0 (56eaea8)&\#1440-2&\textbf{User after free}&\no&gf\_isom\_box\_del&Fixed&Pending\\\cline{3-9}

&&0.8.0 (56eaea8)&\#1440-3&\textbf{User after free}&\no&gf\_isom\_box\_del&Fixed&Pending\\\cline{3-9}

&&0.8.0 (5b37b21)&\#1427&\textbf{User after free}&\yes&gf\_m2ts\_process\_pmt&&\\\cline{3-9}

&&0.7.1 (987169b)&\#1263&NULL pointer dereference&\yes&ilst\_item\_Read&Fixed&\\\cline{3-9}

&&0.7.1 (987169b)&\#1264&Heap buffer overflow&\yes&gf\_m2ts\_process\_pmt&Fixed&CVE-2019-20629\\\cline{3-9}

&&0.7.1 (987169b)&\#1265&Invalid read&\yes&gf\_m2ts\_process\_pmt&Fixed&\\\cline{3-9}

&&0.7.1 (987169b)&\#1266&Invalid read&\yes&gf\_m2ts\_process\_pmt&Fixed&\\\cline{3-9}

&&0.7.1 (987169b)&\#1267&NULL pointer dereference&\yes&gf\_m2ts\_process\_pmt&Fixed&\\\cline{3-9}

&&0.7.1 (987169b)&\#1268&Heap buffer overflow&\yes&BS\_ReadByte&Fixed& CVE-2019-20630\\\cline{3-9}

&&0.7.1 (987169b)&\#1270&Invalid read&\yes&gf\_list\_count&Fixed&CVE-2019-20631\\\cline{3-9}

&&0.7.1 (987169b)&\#1271&Invalid read&\yes&gf\_odf\_delete\_descriptor&Fixed&CVE-2019-20632\\\cline{3-9}

&&0.8.0 (5b37b21)&\#1445&Heap buffer overflow&\yes&gf\_bs\_read\_data&Fixed&\\\cline{3-9}

&&0.8.0 (5b37b21)&\#1446&Stack buffer overflow&\yes&gf\_m2ts\_get\_adaptation\_field&Fixed&\\\cline{1-9}

\multirow{3}{*}{GNU patch} & \multirow{3}{*}{7K}
&2.7.6 (76e7758)&\#56683&\textbf{Double free}&\yes&another\_hunk&Confirmed&CVE-2019-20633\\\cline{3-9}

&&2.7.6 (76e7758)&\#56681&Assertion failure&\yes&pch\_swap&Confirmed&\\\cline{3-9}

&&2.7.6 (76e7758)&\#56684&Memory leak&\no&xmalloc&Confirmed&\\\cline{1-9}

\multirow{8}{*}{Perl 5} & \multirow{8}{*}{184K}
&5.31.3 (a3c7756)&\#134324&\textbf{Use after free}&\yes&S\_reg&Confirmed&\\\cline{3-9}

&&5.31.3 (a3c7756)&\#134326&\textbf{Use after free}&\yes&Perl\_regnext&Fixed&\\\cline{3-9}

&&5.31.3 (a3c7756)&\#134329&\textbf{User after free}&\yes&Perl\_regnext&Fixed&\\\cline{3-9}

&&5.31.3 (a3c7756)&\#134322&NULL pointer dereference&\yes&do\_clean\_named\_objs&Confirmed&\\\cline{3-9}

&&5.31.3 (a3c7756)&\#134325&Heap buffer overflow&\yes&S\_reg&Fixed&\\\cline{3-9}

&&5.31.3 (a3c7756)&\#134327&Invalid read&\yes&S\_regmatch&Fixed&\\\cline{3-9}

&&5.31.3 (a3c7756)&\#134328&Invalid read&\yes&S\_regmatch&Fixed&\\\cline{3-9}

&&5.31.3 (45f8e7b)&\#134342&Invalid read&\yes&Perl\_mro\_isa\_changed\_in&Confirmed&\\\cline{1-9}

\multirow{1}{*}{MuPDF} & \multirow{1}{*}{539K}
&1.16.1 (6566de7)&\#702253&\textbf{Use after free}&\no&fz\_drop\_band\_writer&Fixed&\\\cline{1-9}

\multirow{1}{*}{Boolector} & \multirow{1}{*}{79K}
&3.2.1 (3249ae0
)&\#90&NULL pointer dereference&\yes&set\_last\_occurrence\_of\_symbols&Confirmed&\\\cline{1-9}

\multirow{2}{*}{fontforge} & \multirow{2}{*}{578K}
&20200314 (1604c74)&\#4266&\textbf{Use after free}&\yes&SFDGetBitmapChar&&\\\cline{3-9}

&&20200314 (1604c74)&\#4267&NULL pointer dereference&\yes&SFDGetBitmapChar&&\\\cline{1-9}

\end{tabular} 
\end{adjustbox}
\end{table*}}

\end{document}